\documentclass[10pt,letterpaper]{article}
\usepackage[top=0.85in,left=2.75in,footskip=0.75in]{geometry}

\usepackage{amsmath,amssymb}

\usepackage{changepage}

\usepackage[utf8x]{inputenc}

\usepackage{textcomp,marvosym}

\usepackage{cite}

\usepackage{nameref,hyperref}

\usepackage[right]{lineno}

\usepackage{microtype}
\DisableLigatures[f]{encoding = *, family = * }

\usepackage[table]{xcolor}

\usepackage{array}

\usepackage{longtable}
\newcolumntype{+}{!{\vrule width 2pt}}

\newlength\savedwidth

\newcommand\thickhline{\noalign{\global\savedwidth\arrayrulewidth\global\arrayrulewidth 2pt}%
\hline
\noalign{\global\arrayrulewidth\savedwidth}}

\raggedright
\usepackage[aboveskip=1pt,labelfont=bf,labelsep=period,justification=raggedright,singlelinecheck=off]{caption}

\makeatletter
\renewcommand{\@biblabel}[1]{\quad#1.}
\makeatother

\usepackage{lastpage,fancyhdr,graphicx}
\usepackage{epstopdf}
\usepackage{soul}
\fancyheadoffset[L]{2.25in}
\fancyfootoffset[L]{2.25in}
\usepackage{siunitx}

\begin{document}
\vspace*{0.2in}

\begin{flushleft}
{\Large
\textbf\newline{Closed-loop targeted optogenetic stimulation of \textit{C. elegans} populations} 
}

Mochi Liu\textsuperscript{1},
Sandeep Kumar\textsuperscript{2},
Anuj K Sharma\textsuperscript{1},
Andrew M Leifer\textsuperscript{1,2*}
\\
\bigskip
\textbf{1} Department of Physics, Princeton University, USA
\\
\textbf{2} Princeton Neuroscience Institute, Princeton University, USA
\\
\bigskip

* leifer@princeton.edu

\end{flushleft}
\section*{Abstract}
We present a high-throughput  optogenetic illumination system capable of simultaneous closed-loop  light delivery to specified targets in populations of moving \textit{Caenorhabditis elegans}. The instrument addresses three technical challenges: it delivers targeted illumination  to specified regions of the animal's body such as its head or tail; it automatically delivers stimuli triggered upon the animal's behavior; and it achieves high throughput by  targeting many animals simultaneously. The instrument was used to optogenetically probe the animal's behavioral response to competing mechanosensory stimuli in the the anterior and posterior soft touch receptor neurons. Responses  to more than $10^4$ stimulus events from a range of anterior-posterior intensity combinations were measured. The animal's probability of sprinting forward in response to a mechanosensory stimulus depended on both the anterior and posterior  stimulation intensity, while the probability of reversing depended primarily on the posterior stimulation intensity. We also probed the animal's response to mechanosensory stimulation during the onset of turning, a relatively rare behavioral event, by delivering stimuli  automatically when the animal began to turn. Using this closed-loop approach, over $10^3$  stimulus events were delivered during turning onset at a rate of 9.2 events per worm-hour, a greater than 25-fold increase in throughput compared to  previous investigations. These measurements validate with greater statistical power  previous findings that turning acts to gate mechanosensory evoked reversals.  Compared to previous approaches, the current system offers targeted optogenetic stimulation to specific body regions or behaviors with many-fold increases in throughput to better constrain quantitative models of sensorimotor processing.

\section*{Author summary}
Targeting optogenetic stimulation to only specific body regions or during only specific behaviors is useful for dissecting the role of neurons and circuits underlying behavior in small transparent organisms like the nematode \textit{C. elegans}. Existing methods, however, tend to be very low-throughput, which can pose challenges for acquiring sufficiently large datasets  to constrain quantitative models of neural computations. Here we describe a method to  deliver targeted illumination to many animals simultaneously at high throughput. We use this method to probe mechanosensory processing in the worm and we validate previous experiments with many-fold higher throughput,  larger sample sizes and greater statistical power.



\section*{Introduction}
 
How sensory signals are transformed into motor outputs is a fundamental question in systems neuroscience \cite{clark_mapping_2013}.  Optogenetics  \cite{boyden_history_2011, fenno_development_2011} coupled with automated measures of behavior \cite{datta_computational_2019, pereira_quantifying_2020}  has emerged as a useful tool for probing sensorimotor processing, especially in small  model organisms \cite{calhoun_quantifying_2017}.  In optically transparent animals, such as \textit{C. elegans} and \textit{Drosophila}, such optogenetic manipulations can be performed non-invasively  by  illuminating an animal expressing  light gated ion channels \cite{nagel_light_2005, leifer_optogenetics_2012}. Optogenetically perturbing neural activity and observing behavior has been widely used to study specific neural circuits, such as those involved in chemotaxis \cite{kocabas_controlling_2012, hernandez-nunez_reverse-correlation_2015, gepner_computations_2015, schulze_dynamical_2015} (reviewed  for \textit{Drosophila} in \cite{calabrese_search_2015}), olfaction \cite{gordus_feedback_2015}, learning and memory \cite{claridge-chang_writing_2009, cho_parallel_2016},  and locomotion and escape  \cite{wen_proprioceptive_2012, donnelly_monoaminergic_2013, kato_global_2015,  wang_flexible_2020}, to name just a few examples. In \textit{Drosophila}, high throughput optogenetic delivery to behaving animals has been used to screen libraries of neural cell types  and map out previously unknown relations between  neurons and behavior \cite{cande_optogenetic_2018}.

Optogenetic investigations of neural circuits underlying behavior confront three technical challenges: the first is to deliver stimulation targeted  only to the desired neuron or neurons; the second is to deliver the stimulus at the correct time  in order to probe the circuit  in a relevant  state or behavioral context; and the third is to efficiently acquire enough observations of stimulus and response to draw meaningful conclusions. Existing methods  each address some of these challenges, but not all three.  

To  stimulate only desired cells,  the  expression of  optogenetic proteins is typically  restricted to  specific cells or cell-types \cite{boulin_reporter_2006, pfeiffer_tools_2008, jenett_gal4-driver_2012}. If   cell-specific genetic drivers are not readily available, then genetic specificity can be complemented with optical targeting. Patterned light from a projector, for example, can be used to  illuminate only a subset of the cells  expressing the optogenetic protein \cite{guo_optical_2009, wyart_optogenetic_2009}. For targeting  behaving animals, real-time processing is also needed to track the animal and dynamically update the illumination pattern  based on its movements \cite{leifer_optogenetic_2011, stirman_real-time_2011, bath_flymad_2014, shipley_simultaneous_2014, porto_reverse-correlation_2017, dong_toward_2021}. 

Closed-loop approaches are further needed to  deliver a perturbation timed to a specific  behavior. For example,  delivering a perturbation triggered to the swing of an animal's head has informed our understanding of neural mechanisms underlying   thermotaxis \cite{stephens_dimensionality_2008} and chemotaxis \cite{kocabas_controlling_2012}. 
The use of closed-loop stimulation triggered on behavior in \textit{C. elegans} \cite{faumont_image-free_2011, dong_toward_2021}, \textit{Drosophila} \cite{musso_closed-loop_2019} and  mammals \cite{adamantidis_optogenetic_2011, oconnor_neural_2013}
is part of a broader trend in systems neuroscience towards more efficient exploration of  the vast space of possible neural perturbations \cite{clancy_volitional_2014, grosenick_closed-loop_2015}, especially during ethnologically relevant naturalistic behaviors \cite{krakauer_neuroscience_2017}.

Targeted and closed-loop illumination systems probe one animal at a time  \cite{leifer_optogenetic_2011, stirman_high-throughput_2010, kocabas_controlling_2012, lee_compressed_2019, dong_toward_2021}, or at most  two  \cite{wu_optogenetic_2014}. This low throughput poses challenges for acquiring  datasets with enough statistical power to constrain quantitative models of neural computation. To increase throughput, a  separate  body of work simultaneously measures behavior of populations of many animals in an arena, in order to amass thousands of animal-hours of recordings  \cite{kerr_imaging_2006, branson_high-throughput_2009,  gershow_controlling_2012, husson_keeping_2012}. Delivering spatially uniform  optogenetic perturbations to such populations has helped constrain quantitative models of sensorimotor processing of chemotaxis and mechanosensation \cite{gepner_computations_2015, hernandez-nunez_reverse-correlation_2015, liu_temporal_2018}. Because  the entire body of every animal is illuminated,  this approach relies entirely on genetics for targeting. Recent work has used  stochastic spatially varying illumination patterns in open-loop from a projector \cite{deangelis_spatiotemporally_2020}  or cellphone \cite{meloni_controlling_2020}  to  probe the spatial dependence of optogenetic stimulation. But because these methods are open loop they cannot  target stimulation specifically to any animal or body part. Instead they  rely on after-the fact inspection of where their stimuli landed, decreasing throughput.

Here we demonstrate a closed-loop real-time targeting system for \textit{C. elegans} that tackles all three challenges: targeting, closed-loop triggering and throughput. The system  delivers targeted illumination to specified parts of the animal's body,  stimulus delivery can be be triggered automatically on behavior, and the system achieves a high throughput by tracking and  independently delivering targeted stimuli to populations of animals simultaneously. We apply this system to the \textit{C. elegans} mechanosensory circuit \cite{chalfie_developmental_1981, chalfie_neural_1985} to characterize how competing stimuli  in anterior and posterior mechanosensory neurons are integrated by downstream circuity. We also revisit  our prior observation  that turning behavior gates mechanosensory evoked reversals \cite{liu_temporal_2018}. We deliver closed-loop stimulation  triggered to the onset of turning to obtain a dataset with  two-orders-of-magnitude more stimulus-events compared to that investigation. We use these measurements to  validate our initial observation with greater statistical power.

\section*{Results}

\begin{figure}[!htbp]
	\begin{center}
		\includegraphics[width=0.8\textwidth]{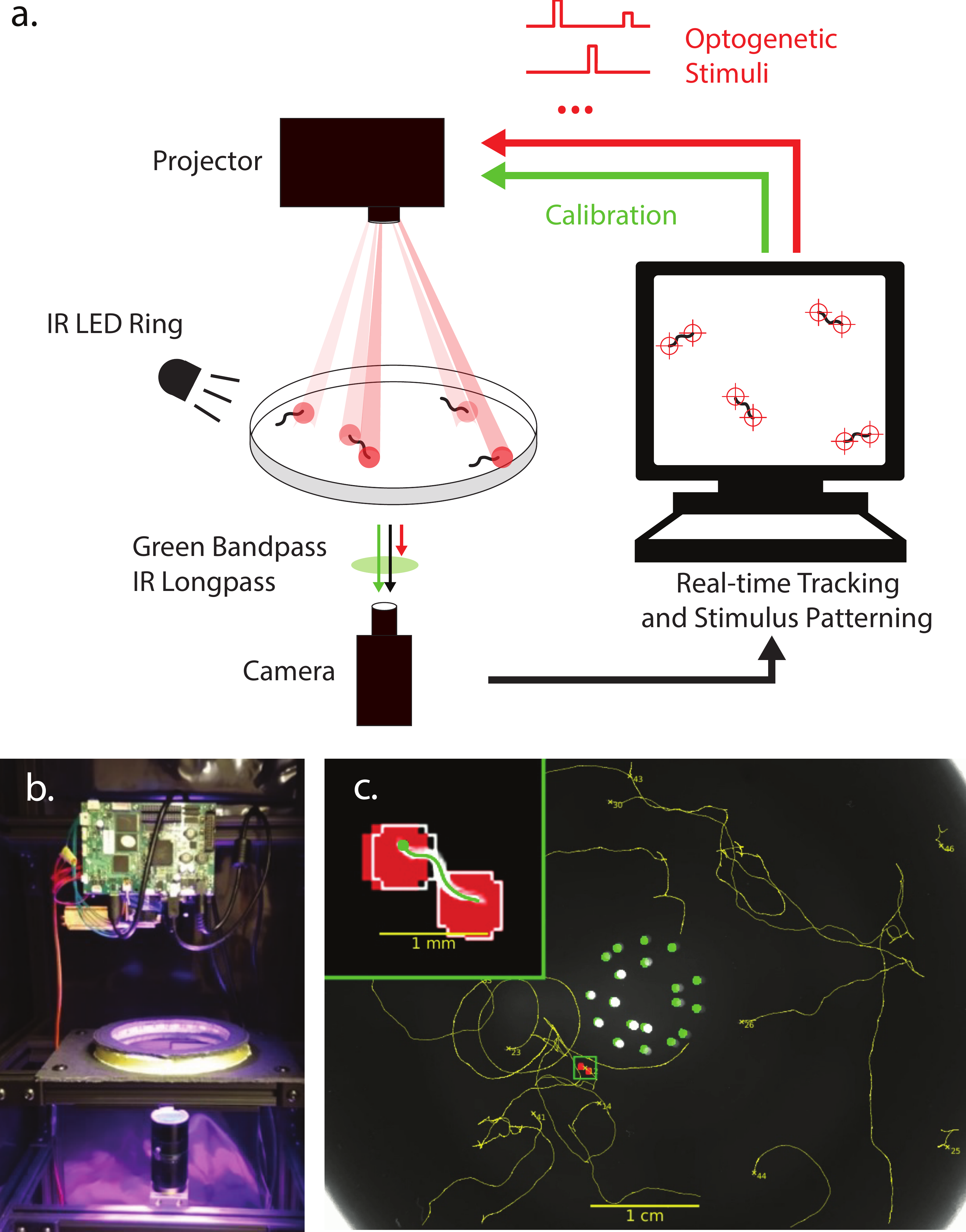}
		\caption{{\bf Closed-loop  targeted optogenetic delivery system.} a) Schematic of system. A projector simultaneously delivers  patterned targeted optogenetic stimulation to multiple worms on an agar plate in real-time. b) Photograph of instrument. c) Image from experiment shows 
		animals expressing Chrimson in touch receptor neurons (AML470) as they crawl on agar. Corresponding video is shown in \nameref{vid:arena}.  Tracked paths are shown in yellow. Green and white dots in the center relate to a visual time-stamping system and are excluded from analysis.  Inset shows details of an animal receiving optogenetic stimulation targeted to its head and tail (0.5mm diameter stimuli).   The two white circle in the inset show the targeted location of the stimulus. Red shading shows area where stimulus was  delivered. }
		\label{fig:instrument}
	\end{center}
\end{figure}

We developed a closed-loop targeted delivery system to illuminate specified regions in populations of \textit{C. elegans} expressing the optogenetic protein Chrimson as the animals crawled on agar in a 9 cm dish. The system used red light (peak 630 nm) from a custom color projector made of an optical engine (Anhua M5NP) driven by a separate control board, described in methods. Animal behavior was simultaneously  recorded  from a CMOS camera, Figure \ref{fig:instrument}a,b.   Dark-field illumination from a ring of infrared LEDs (peak emission at 850 nm) was used to observe animal behavior because this avoided exciting Chrimson. Optical filters allowed the infrared illuminated behavior images to reach the camera, but blocked   red or blue light coming from the projector. Green light from the projector was used to  present  visual timestamps and other spatiotemoporal calibration information  to the camera.   Custom real-time computer vision software  monitored the pose and behavior of each animal and  generated patterned illumination restricted to only specified regions of the animal, such as its head or tail, optionally triggered by the animal's behavior, Figure \ref{fig:instrument}c and \nameref{vid:arena}.

\subsection*{Anterior and posterior mechanosensory response}
The  mechanosensory  response to  anterior or posterior stimulation has been probed extensively by softly touching the animal with a hair \cite{chalfie_developmental_1981, chiba_developmental_1990, chalfie_assaying_2014},   tapping the substrate on which the animal crawls \cite{chalfie_developmental_1981, wicks_integration_1995},
 automatically applying forces via micro-electromechanical force probes \cite{petzold_mems-based_2013, mazzochette_real_2013, mazzochette_tactile_2018} or applying forces via a microfluidic  device \cite{mcclanahan_comparing_2017}. These approaches have identified dozens of genes related to touch sensation \cite{chalfie_neurosensory_2009}.
 
Mechanosensory responses have also been probed optogenetically by expressing light-gated ion channels in the six soft-touch mechanosenory neurons  ALML, ALMR, AVM, PVM, PLML and PLMR \cite{nagel_light_2005}. Targeted illumination experiments showed that optogenetically stimulating posterior touch neurons PLML and PLMR was sufficient to elicit the animal to speed up or sprint and that stimulating   anterior soft touch neurons (ALML, ALMR) or AVM was sufficient to induce the  animal to move backwards in a reversal behavior   \cite{leifer_optogenetic_2011, stirman_real-time_2011, porto_reverse-correlation_2017}. Those experiments were conducted one worm-at-a time, to accommodate  the patterned illumination systems that were needed to restrict illumination to only a portion of the animal.   Here we revisit these  experiments, this time by probing a population of animals simultaneously to achieve higher sample sizes. 

\begin{figure}[htbp]
	\begin{adjustwidth}{-2.25in}{0in} 
	\begin{center}
		\includegraphics[width=1.37\textwidth]{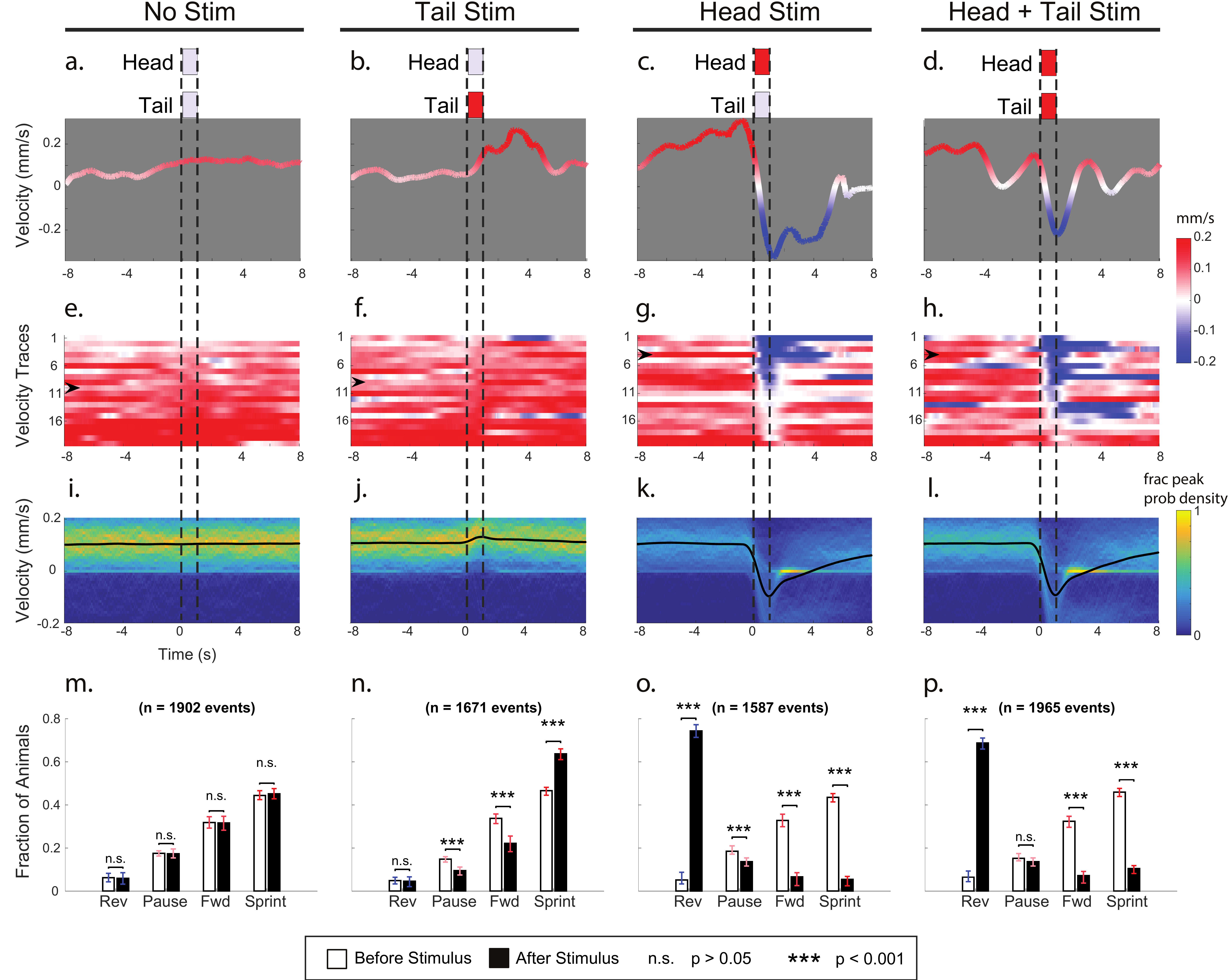}
		\caption{{\bf Stimulation of anterior mechanosensory neurons evokes reverse behavior; stimulation of posterior mechanosensory neurons  evokes sprints.} Targeted stimulation is delivered to  worms expressing Chrimson in their soft touch mechanosensory neurons (strain AML470). \textbf{a-d)} Representative velocity trace from single stimulus event.  Dashed lines indicate timing of stimulation  (1 s,  0.5mm diameter circular illumination pattern centered at the tip of the animal's head and/or tail,  red light intensity of 80 uW/mm\textsuperscript{2}). \textbf{e-h)} 20 randomly sampled velocity traces  for each stimulus condition are shown, sorted by  mean velocity immediately after stimulation. Same as  in Supplementary Videos S2-S5. Arrow indicates each representative trace shown in \textbf{a-d}. \textbf{i-l)} Probability density of velocities for each stimulus condition. Mean velocity is shown as black line. $n>1,500$ stimulus-events per condition.  \textbf{m-p)} The fraction of animals inhabiting each of four velocity-defined behavioral states  is shown for the time point 2 s before stimulation onset and  immediately after stimulation. Cutoffs for behavior states are shown in \nameref{fig:velocitybehaviors}.  P-values calculated using Wilcoxon rank-sum test. Error bars represents 95\% confidence intervals estimated using 1000 bootstraps.}
		\label{fig:apresponse}
	\end{center}
	\end{adjustwidth}
\end{figure}

We delivered 1 s of red light  illumination  to the  head, tail, or both, of transgenic animals expressing  the light-gated ion channel Chrimson in the six soft touch mechanosensory neurons (500 um diameter stimulus region, 30 s inter-stimulus interval.)  Multiple combinations of illumination intensities were used, but here only the 80 uW/mm$^2$ intensity stimuli are considered, Figure~\ref{fig:apresponse}. The system independently tracked or stimulated an average of $12\pm 10$ worms on a plate simultaneously (mean $\pm$ standard deviation), \nameref{tab:recordings}. Across four days, 95 plates were imaged using  four identical instruments running in parallel, for a total of 31,761 stimulation-events across all stimulus intensity combinations. Of those, there were 7,125 stimulus-events with an illumination intensity of  80 uW/mm$^2$,  at least 1,500 stimulus events  each of the following conditions: head illumination, tail illumination, both, or neither. For comparison, \cite{leifer_optogenetic_2011} used 14 stimulus events, a two-order of magnitude difference in sample size.

Consistent with prior reports, activating posterior mechanosensory neurons by delivering light to the tail resulted in an increase in sprints and an overall increase in average velocity (Figure \ref{fig:apresponse} and \nameref{vid:tail}). Similarly, activating anterior mechanosensory neurons by delivering light to the head resulted in an increase in reversal events and a decrease in average velocity (\nameref{vid:head}). Simultaneous stimulation to head and tail resulted an increase in reversals (\nameref{vid:headandtail}). Animals were classified as performing forward locomotion, pause, sprint or reversals, based on their velocity,  \nameref{fig:velocitybehaviors}.  The animal showed little response to control conditions in which no light was delivered (Figure \ref{fig:apresponse}a,e,i,m and \nameref{vid:control}), or in which the necessary co-factor all-trans retinal was withheld, \nameref{fig:anteriorposteriornoret}.

We also   stimulated a different strain of animals nominally of the same genotype that we had used previously, AML67,  \cite{liu_temporal_2018}. That strain had been constructed with a   higher DNA plasmid concentration during microinjection (40 ng/ul for AML67, compared to  10 ng/ul for AML470 used above). The AML67 strain  behaved the same on this instrument in response to simultaneous anterior and posterior stimulation as it had in \cite{liu_temporal_2018}. It also exhibited reversals in response to anterior illumination, as expected, but surprisingly, it  did not exhibit sprint behaviors in response to posterior illumination as AML470 did (\nameref{fig:AML67anteriorposterior}). We suspect that some aspect resulting from the higher DNA plasmid injection concentration caused animals of this strain to behave differently during sprint behaviors.

\subsection*{Integration of conflicting anterior and posterior mechanosensory signals}
Mechanosensory neurons  act as inputs to downstream interneurons and motor neurons that translate mechanosensory signals into a behavioral response \cite{chalfie_neural_1985, wicks_integration_1995, gray_circuit_2005, alkema_tyramine_2005, wang_flexible_2020}. A growing body of evidence suggests that  downstream circuitry relies on the magnitude of signals in both the anterior and posterior mechanosensory neurons to determine the behavioral response.   For example, a plate tap activates both anterior and posterior mechanosensory neurons and usually elicits a reversal response \cite{chiba_developmental_1990, wicks_integration_1995, swierczek_high-throughput_2011, liu_temporal_2018}. But the animal's response to tap can be biased towards one response or another by selectively killing  specific touch neurons via laser ablation  \cite{wicks_integration_1995}. For example, if   ALMR alone is ablated, the animal is more balanced in its response and is just as likely to respond to a tap by reversing as it is by sprinting. If both  ALML and ALMR are ablated  the  animal will then sprint  the majority of the time \cite{wicks_integration_1995}.
Competing anterior-posterior optogenetic stimulation of the mechanosensory neurons also influences the animal's  behavioral response. For example,  a higher intensity optogenetic stimulus to the anterior touch neurons is needed to evoke a reversal when  the posterior touch neurons are also stimulated, compared to when no posterior stimulus is present \cite{stirman_real-time_2011}. 

\begin{figure}[htbp]
	\begin{center}
		\includegraphics[width=\textwidth]{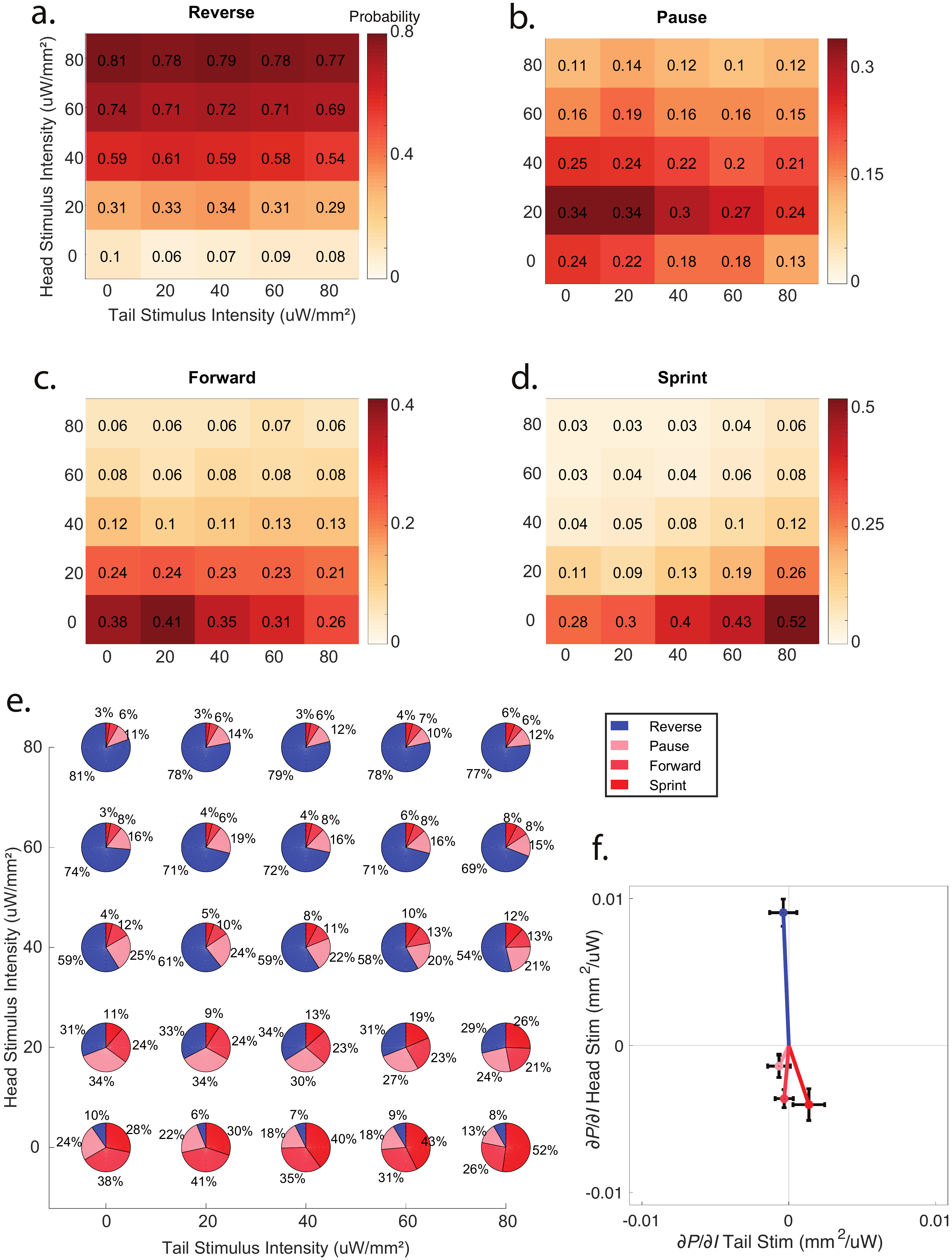}
		\caption{{\bf Behavioral response to competing  stimulation of anterior and posterior mechanosensory neurons.}  Various combinations of light intensity  was  delivered to the head and tail of worms expressing Chrimson in soft-touch mechanosensory neurons (strain AML470, n= 31,761 stimulus-events total, supserset of data shown in Figure \ref{fig:apresponse} ). a.) Probability of reverse b.) pause c.) forward and d.) sprint behaviors are shown individually e) and all together as pie charts. f.) The gradient of the plane-of-best-fit is shown as a vector for each behavior. Fitting was performed using methods of least squares. Error-bars are 95\% confidence intervals.}
		\label{fig:heatmap}
	\end{center}
\end{figure}

To systematically characterize how anterior and posterior mechanosensory signals are integrated, we inspected the animal's probability of exhibiting reverse, forward, pause or sprint behavior in response to 25 different combinations of head and tail light intensity stimuli, Figure \ref{fig:heatmap}a-d. This data is a superset of that shown in Figure \ref{fig:apresponse}. Here 31,761 stimulus-events are shown, corresponding to all 25 different conditions. Behavior is defined such that the animal always occupies one of four states: reverse, pause, forward or sprint, so that for a given stimulus condition, the four probabilities necessarily sum to one, Figure \ref{fig:heatmap}e.  

To explore the dependence of the probability of evoking a given behavior on the anterior and posterior stimulus intensity, we  fit planes to the probability surfaces and computed the gradient of the resulting plane, Figure \ref{fig:heatmap},f.  Fitting was performed using the method of least squares. The head and tail components of the gradient provide a  succinct estimate of how the probability depends on either  head or tail stimulus illumination intensity. For example, the probability of reversal  depends strongly on the head stimulus intensity, as indicated by the large positive head component of the gradient. The probability of reversal also depends slightly on the tail stimulus intensity, consistent with \cite{stirman_real-time_2011}, but we note this dependence was  small and that the 95\% confidence intervals of this dependence spanned the zero value.   Of the four behaviors tested, only  sprint behavior depended on both head and tail stimulation intensity such that the 95\% confidence intervals of both components of their gradient excluded the value zero. Sprints occurred with highest probability at the highest tail illumination intensity when head illumination was zero. As head illumination  increased, the probability of a sprint rapidly decreased.  One interpretation of these measurements is that  head  induced reversals are less likely to be counteracted by a  tail stimulation, than tail  induced sprints are to be counteracted by  head stimulation. 

Taken together, we conclude that anterior and posterior mechanosensory signals are combined in a nontrivial way to determine the relative likelihood of different behavioral responses. For example, the animal's decision to reverse in response to mechanosensory stimulation is  primarily dependent on signals in anterior mechanosensory neurons with only minimal dependence on posterior mechanosensory neurons. In contrast, the animal's decision to sprint is  influenced more evenly by signals in both sets of neurons.  This places constraints on any quantitative models of  sensory integration performed by downstream circuitry.

\subsection*{Behavior-triggered stimulation increases throughput when investigating rare behaviors}
\textit{C. elegans'} response to mechanosensory stimulation depends on its behavioral context. The probability that a tap or optogenetic mechanosensory stimulation evokes a reversal is higher when the stimulus occurs as the animal moves forward compared to  if the stimulus occurs when the animal is in the midst of a turn, suggesting that the nervous system gates mechanosensory evoked responses during turns \cite{liu_temporal_2018}. This was first observed in  open-loop experiments in which the animal was stimulated irrespective of its behavior. Those experiments relied on identifying, after-the-fact,  stimulus-events that arrived by chance during turning. Because turning events are brief and occur infrequently,  it can be challenging to observe sufficient numbers of stimuli events delivered during turn onset using such an open-loop approach. For example, in that work only 15 optogenetic stimulus-events landed during turns. 
The animal's spontaneous rate of turn initiation varies with the presence of food and other  environmental conditions, but it has been reported to be approximately 0.03 Hz \cite{pierce-shimomura_fundamental_1999}.  

To obtain higher efficiency and throughput at probing the animal's response to stimulation during turning, we delivered closed-loop stimulation automatically triggered on the onset of the animal's turning behavior.  
Full-body red  light illumination (1.5 mm diameter) was automatically delivered to animals expressing Chrimson in the soft-touch mechanosensory neurons (strain AML67, same as  in \cite{liu_temporal_2018})  when the real-time tracking software detected that the animal entered a turn. 
Turn onset was detected by monitoring the ellipse ratio of a binary image of the animal, as described in methods.  
A refractory period of 30 s was imposed to  prevent the same turning event from triggering multiple stimuli and served to set a minimum inter-stimulus interval. 
47 plates of animals were recorded  for 30 mins each over three days, and on average, the system simultaneously tracked or stimulated 44.5$\pm$20 worms on a plate at any given time, \nameref{tab:recordings}.  
Three different stimulus intensities (0.5, 40 and 80 uW/mm$^2$) and three different stimulus durations  (1, 3 and 5 s) were explored, totaling  39,477 turn-triggered stimuli events, of which on post-processing analysis 9,774 or 24.8\% passed our more stringent inclusion criteria for  turn onset, worm validity and track stability, Table \ref{tab:enrichmentcomparison}.
To compare the closed- and open-loop approaches,  29 additional plates were stimulated in open loop over the same three day period.

\begin{figure}[!h]
	\begin{center}
		\includegraphics[width=\textwidth]{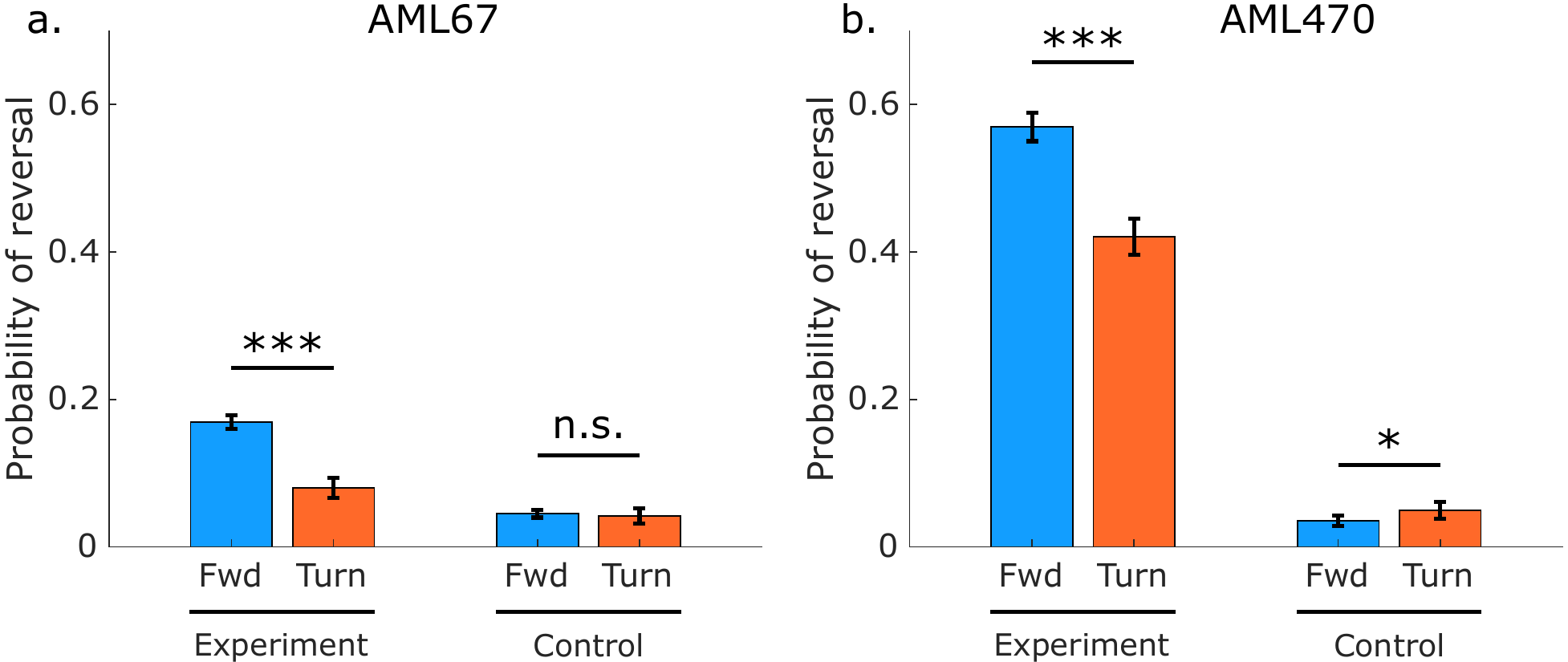}
		\caption{{\bf Probability of reversing in response to a mechanosensory stimulus is higher for stimuli that arrive during forward locomotion than for stimuli that arrive during turning.} Response to  whole body optogenetic illumination of soft-touch mechanosensory neurons  is shown for stimuli that arrive during either forward or turning behaviors for two strains of nominally identical genotypes, a) AML67 and b) AML470.    Stimuli delivered during turns are   from closed-loop optogenetic experiments, while stimuli delivered during forward locomotion are from  open-loop experiments. 3 s of 80 uW/mm$^2$ illumination was delivered in the experiment condition. Only 0.5 uW/mm$^2$ was delivered for control condition.  Error bars are 95 percent confidence interval calculated via 10,000 bootstraps. Z-test was used to calculate significance. *** indicates $p<0.001$. $p$ value for AML67 control group is 0.549. $p$ value for AML470 control group is 0.026. The number of stimulus events for each condition (from left-most bar to right-most bar) for AML67 are: 5,968, 1,551, 5,971, 1,448; for AML470: 2,501, 1,543, 2,676, 1,438. }
		\label{fig:turning}
	\end{center}
\end{figure}

We compared the probability of reversing in response to  closed-loop stimuli delivered during turn onset against the probability of reversing in response to open-loop stimuli  delivered during forward locomotion, Figure \ref{fig:turning}a. For this analysis we  considered only  stimuli of 3 s duration and either  80 uW/mm$^2$  or 0.5 uW/mm$^2$ (control) illumination intensity, of which  2,999 stimulus-events were delivered during turn onset. Consistent with previous reports, the animal was significantly more likely to initiate a reversal in response to stimuli delivered during forward locomotion than during turning. We repeated this experiment  in strain AML470. That strain was also statistically significantly more likely to reverse in response to stimuli delivered during forward locomotion than during turning, although interestingly the effect was less striking in this strain compared to AML67 even though animals were overall more responsive. By using a  high throughput closed-loop approach  we confirmed previous findings with larger sample size ($10^3$ events compared to 15), and revealed subtle differences between two different  strains with nominally identical genotypes.

Both throughput and efficiency are relevant for studying stimulus response during turning. Throughput refers to the number of stimuli delivered during turns per time. High throughput  is needed to generate a  large enough sample size in a reasonable enough amount  time  to draw statistically significant conclusions.  Efficiency, or yield, refers to the fraction of delivered stimuli that were indeed delivered during turns. A high efficiency, or yield, is desired  to avoid  unwanted stimulation of the animal which can lead to unnecessary habituation.

We compared the throughput and efficiency of  stimulating during turn onset with closed-loop stimulation  to an open-loop approach  on the same instrument using our same analysis pipeline and inclusion criteria, Table \ref{tab:enrichmentcomparison}. Again we considered only stimuli delivered within a small 0.33 s window corresponding to our definition of the onset of turns, and applied in post-processing the same stringent inclusion criteria  to both  open-loop and closed-loop  stimuli. Closed loop-stimulation achieved a throughput  of 9.2 turn-onset associated stimulation events per worm hour,  an order of magnitude greater than the 0.5 events per worm hour in open loop stimulation. Crucially, closed-loop stimulation  was also more  efficient,  achieving a yield of 24.8\%,  nearly two orders of magnitude higher than the  0.4\% open-loop yield. We reach similar conclusions by comparing to previous open-loop optogenetic experiments from \cite{liu_temporal_2018} that had a longer inter-stimulus interval. Taken together, by delivering stimuli triggered on turns in a closed-loop fashion, we achieved higher throughput and efficiency than open-loop approaches.

\begin{table}[!ht]
	\begin{adjustwidth}{-2.25in}{0in} 
		\centering
		\caption{ 
			{\bf Comparison of open- and closed-loop approaches for studying the animal's response to stimulation during a turn.} Whole-body illumination experiments using  AML67 are shown. This is a superset of the data shown in Figure \ref{fig:turning} and includes a variety of stimulus intensities and stimulus durations. Compared to an open-loop approach, closed-loop turn-triggered stimulation provides  higher throughput and higher yield. $^*$Note we report an effective throughput for  experiments in \cite{liu_temporal_2018} to account for discrepencies in how different stimulus intensities are reported in that work (reported cumulative worm-hours include 6 different stimuli intensities, while only  one  intensity is considered in the reported number of stimulus events).}
		\begin{tabular}{|l|l|l|l|l|l|l|l|l|}
			\hline
			 Ref. & Experiment Type & Plates & \begin{tabular}[c]{@{}l@{}}Cum. Recording \\ Duration \\(Worm-hr)\end{tabular}  &  \begin{tabular}[c]{@{}l@{}}ISI \\(s)\end{tabular} & \begin{tabular}[c]{@{}l@{}}All Stim \\Events\end{tabular}  & \begin{tabular}[c]{@{}l@{}}Turn-Associated \\ Stim Events\end{tabular} 
			& Yield
			& \begin{tabular}[c]{@{}l@{}}Throughput \\ (Turn-Associated\\ Stim  Events/\\Worm-hr)\end{tabular}  \\ \thickhline

			\begin{tabular}[c]{@{}l@{}}This\\work\end{tabular} & \begin{tabular}[c]{@{}l@{}}Closed-loop\\optogenetic\end{tabular} & 47 & 1,060 & $>30$ & 39,477 &  9,774  & 24.8\% & 9.2   \\ \hline
			
			\begin{tabular}[c]{@{}l@{}}This\\work\end{tabular} & \begin{tabular}[c]{@{}l@{}}Open-loop\\optogenetic\end{tabular} & 29 & 633 & 30 & 70,699 &  299  & 0.4\% & 0.5   
			\\ \thickhline
			
			\cite{liu_temporal_2018} & \begin{tabular}[c]{@{}l@{}}Open-loop\\optogenetic\end{tabular} & 12 & 260 & 60 &  2,487 &  15  & 0.6\% & 0.35$^*$    
			\\ \thickhline

		\end{tabular}
		\label{tab:enrichmentcomparison}
	\end{adjustwidth}
\end{table}

\subsection*{Further characterization of the instrument}
\begin{figure}[!hptb]
	\includegraphics[width=.95\textwidth]{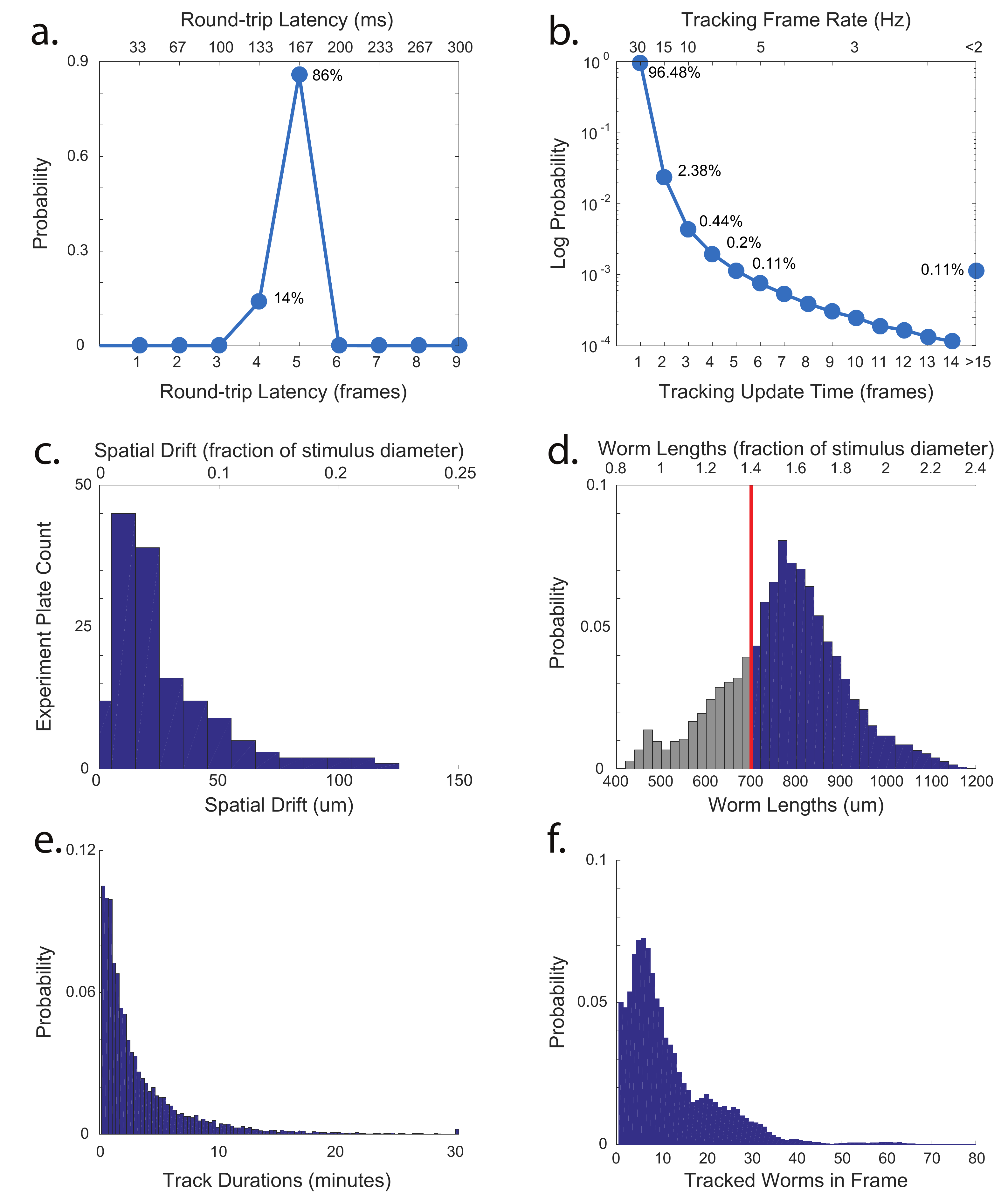}
\caption{{\bf Characterization of key performance metrics.} Performance is evaluated for the set of experiments displayed in Figure \ref{fig:heatmap}.  \textbf{a)} Round-trip latency is the elapsed time between  camera exposure and stimulus delivery in response to that acquired image, as determined by visual time stamps drawn by the projector.  \textbf{b)} The probability distribution of update times for tracking worms is plotted on a log axis. Over 96\% of frames do experience the full 30 Hz frame rate. Around 1 in 1000 frames have a tracking rate of less than 2 Hz. \textbf{c)} Histogram of camera-to-projector spatial drift between start and end of 30 min recording for each of 151 plates. \textbf{d)} Probability distribution  of worm lengths for all tracks. Only worms with lengths 700 um and above are included for behavioral analysis. \textbf{e)} Probability distribution of the duration of tracks in the dataset. \textbf{f)} The probability distribution of the number of tracked worms at any given time. The mean and standard deviation for this set of recordings is $12\pm10$. Mean for other recordings is listed in \nameref{tab:recordings}. }
\label{fig:characterization}
\end{figure}

We sought to further characterize the system to better understand its capabilities and limitations,  Figure \ref{fig:characterization}. We quantified round-trip latency, prevalence of dropped frames, the spatial drift between camera and projector,  and other key parameters that impact performance or  resolution based on an analysis of the recordings in Figure \ref{fig:heatmap} and \nameref{fig:anteriorposteriornoret}.  

The round-trip latency  between an animal's movement and a corresponding change in illumination output based on that movement is a key determinant of system performance. If the latency is too high compared to the animal's motion, the stimulus will arrive off-target. Round-trip latency is the cumulative delay from many steps including: exposing the camera, transferring camera images to memory,  image processing, generating an illumination pattern,  transferring the pattern to the projector and then ultimately adjusting micromirrors inside the projector to illuminate targeted regions of the animal. We constantly measured round-trip latency in real time by   projecting a frame stamp  visible to the camera using the projector's green channel (green and  dots arranged in a circle visible in the  center of Figure \ref{fig:instrument}c. They sometimes appear white in this visualization due to saturation.). The round-trip latency is the elapsed time   between projecting a given frame stamp, and generating a new illumination pattern in response to the camera image  containing that frame stamp. Median round-trip latency was 167 ms, Figure \ref{fig:characterization}a. For a worm moving at a typical center-of-mass velocity of order 200 um/s this corresponds to a roughly 50 um bound on spatial resolution. This is more than sufficient for the 500 um diameter head-tail illuminations used here. But the latency for this system is notably  longer than  single-worm targeted illumination systems (e.g. 29 ms for \cite{shipley_simultaneous_2014}) and suggests that the current system is unlikely to be well-suited to  target the dorsal versus ventral side of the animal, for example. 

When a real-time tracking systems fails to keep pace with a stream of incoming images it may drop the processing of frames,   resulting in a lower effective frame rate.   The system records the acquisition time stamp of each camera frame  that was successfully processed by the real-time tracker. Later, the system detects any frames that were dropped by inspecting gaps in these time stamps. 96.5\% of the time the system achieved 30 Hz with no frame drops, Figure \ref{fig:characterization}c. 99.9\% of the time the system achieved a framerate above 2Hz, which we estimate is roughly the limit for resolving the head versus tail for a 1 mm worm moving at 200 um/s. When frames were dropped, the majority of time only one frame in a row was dropped  (2.38\% of all frames).

Spatial resolution relies on the alignment between the projector and the camera, but this  alignment can shift during the course of a recording due to projector heating or other effects.  We quantified the drift in alignment between projector and camera by comparing a calibration pattern projected onto the agar in the green channel at the beginning and end of each 30 min plate recording. For the majority of recordings, spatial drift was  small, corresponding to less than 25 um, or less than 5\% of the diameter of the head-tail stimulus, Figure \ref{fig:characterization}c.  Finally, we also  report the number of simultaneously tracked worms, the duration for which they were tracked and the worms' size for comparison, Figure \ref{fig:characterization}d-f.

\section*{Discussion and Conclusions}
The approach here dramatically improves throughput in two ways compared to previous methods.  First, this work extends targeted illumination to many worms in parallel, providing an order of magnitude higher throughput per plate compared to previous single-worm methods capable of delivering optogenetic illumination targeted to different regions of the body \cite{leifer_optogenetic_2011, stirman_real-time_2011, kocabas_controlling_2012}. Second, the method  enables automatic stimulus delivery triggered on a behavior. 
For studying  stimulus response during rare behaviors, like turns, this closed-loop approach provides higher throughput and efficiency compared to previous open-loop methods \cite{liu_temporal_2018}. 

To achieve simultaneous independent targeting of many animals and tracking of behavior, we  developed new algorithms, such as  real-time centerline tracking algorithms, improved existing ones using parallelization and also  leveraged advances in computing power. For example, we  took advantage of the availability of powerful fast many-core CPUs,  GPUs,  multi-terabyte solid-state drives, and low-latency USB3 cameras.  

To attain such high-throughput, the system also sacrifices some spatial resolution compared to previous single-worm approaches \cite{leifer_optogenetic_2011, stirman_high-throughput_2010, kocabas_controlling_2012}.  For example, the round-trip latency and observed drift in calibration  places a roughly 100 um floor on our spatial resolution, which makes the system ill-suited for resolving individual neurons located close to one another.   Nonetheless,  this resolution is more than sufficient to selectively illuminate the head or tail of adult \textit{C. elegans}, which allows for new types investigations. For example, we used the instrument to systematically probe anterior-posterior integration of mechanosensory signals for a range of competing stimuli intensities, delivering over $3.1\times10^4$ stimulus-events in total.   The sample size needed for such an experiment would be impractical with single-worm targeted illumination methods. And current genetic labeling approaches preclude this experiment from being conducted with non-targeted whole field-of-view illumination setups, such as in \cite{liu_temporal_2018}. 

Our measurements  suggest  that the worms' behavioral response to competing mechanosensory stimuli depends on integrating anterior and posterior mechanosensory signals in a non-trivial way. The probability of a sprint is influenced roughly evenly by signals in both anterior and posterior mechanosensory neurons, while the probability of reversing is primarily influenced by the anterior mechanosensory neurons.  Overall, head stimuli that would induce reversals are less likely to be counteracted by a  tail stimulation, than tail  induced sprints are to be counteracted by  head stimulation. The \textit{C. elegans} response to anterior mechanosensory stimuli is an important part of the escape response \cite{pirri_neuroethology_2012} and helps the animal avoid predation by nematophagous fungi \cite{maguire_c._2011}. It is possible that the relative difficulty in disrupting head induced reversals compared to sprints reflects  the relative importance of the role of the reversal in this escape behavior. 

Here we used red  illumination to excite Chrimson, but we note that the system can be trivially extended to  independently deliver simultaneous red and blue light illumination  \cite{stirman_real-time_2011}, for example to  independently activate two different opsins such as the excitatory red opsin Chrimson and the inhibitory blue opsin gtACR2 \cite{govorunova_natural_2015, vierock_bipoles_2021}. Like other targeted illumination systems before it \cite{leifer_optogenetic_2011, stirman_real-time_2011}, this system is not capable of  targeting  regions within the body when the animal touches itself, as often occurs during turning, or when coiling \cite{croll_behavoural_1975, croll_components_1975}. This still permits probing the animal's response to mechanosensory stimulation during turns because we were interested in  whole-animal stimulation for those specific experiments, rather than targeting the  head or tail.   We note that our post-processing analysis   does resolve the  animal's centerline even during self-touching \cite{liu_temporal_2018},  but that method is not currently suitable for real-time processing. 

We investigated the response to stimulus during turning by delivering closed-loop stimuli automatically triggered on the turn.
We achieved a more than 25-fold increase in  throughput  compared to a previous investigation  \cite{liu_temporal_2018} and  similar order-of-magnitude increase compared to an open-loop approach implemented on the same instrument with the same analysis pipeline and inclusion criteria. The  closed-loop functionality can be easily triggered on sprints or pauses or, in principal, even on extended motifs like an escape response. This high-throughput triggering capability may be useful for searching for long-lived behavior states, probing the hierarchical organizations of behavior \cite{kaplan_nested_2020}, or exploring other instances of context-dependent sensory processing \cite{liu_temporal_2018}.

\section*{Materials and methods}
\subsection*{Strains}
Two strains were used in this work, AML67 and AML470, Table \ref{table:methods:strains}. A list of strains cross-referenced by figure is shown in \nameref{tab:recordings}. Both strains expressed  the light gated ion channel Chrimson and a fluorescent reporter mCherry under the control of a \textit{mec-4} promoter, and differed mainly by the concentration of the Chrimson-containing plasmid used for injection. AML67 was injected with 40 ng/ul of the Chrimson-containing plasmid  while AML470 was injected with 10 ng/ul.   Specifically, to generate AML470,  a plasmid mix containing 10 ng/ul of pAL::pmec-4::Chrimson::SL2::mCherry::unc-54 (RRID:Addgene\_107745) and 100 ng/ul of pCFJ68 unc-122::GFP (RRID:Addgene\_19325) were injected into CZ20310  \cite{noma_rapid_2018} and then integrated via UV irradiation. Experiments were conducted with AML470 strains  prior to outcrossing.

\begin{table}
\begin{adjustwidth}{-2.35in}{0in}
\centering
\caption{
{\bf Strains  used.}} 
\begin{tabular}{|c|p{4.5cm}|p{7.3cm}|p{2.9cm}|p{1cm}|}
\hline
Strain & RRID & Genotype & Notes & Ref   \\
\hline
AML67 & RRID:WB-STRAIN:WBStrain00000193 &wtfIs46[pmec-4::Chrimson::SL2::mCherry::unc-54 40ng/ul] & 40 ng Chrimson injection & \cite{liu_temporal_2018}\\ \hline
AML470  &  & juSi164 unc-119(ed3) III; wtfIs458 [mec-4::Chrimson4.2::SL2::mCherry::unc-54 10 ng/ul + unc-122::GFP 100 ng/ul] & 10 ng Chrimson injection & This work\\ \hline
\end{tabular}
\label{table:methods:strains}
\end{adjustwidth}
\end{table}

\subsection*{Instrument}
\subsubsection*{Hardware}
A CMOS camera (acA4112-30um, Basler) captured images of worms crawling on a 9 cm diameter agar plate at 30 frames per second, illuminated by a ring of 850 nm infrared LEDs, all housed in a custom  cabinet made of 1 inch aluminum extrusions. To illuminate the worm, a custom  projector was built by combining a commercial DMD-based light engine (Anhua M5NP, containing a Texas Instruments DLP4500) with a Texas Instrument evaluation control board (DLPLCR4500EVM). The light engine contained red, green and blue LEDs with peaks at 630 nm, 540 nm and 460 nm, respectively (\nameref{fig:spectra}). The projector   cycles sequentially through patterns illuminated by red, green and then blue illumination once per  cycle  at up to 60 Hz and further modulates the perceived illumination intensity for each pixel within each color by fluttering  individual mirrors with varying duty-cycles at 2.8 kHz.   The system produced a small image (9 cm wide) from a short distance away (15 cm) such that a single element of the DMD projects light onto  a roughly 85 um$^2$ region of agar. 

A light engine  driven by a separate evaluation board was chosen instead of an all-in-one off-the-shelf projector  because the API provided by the evaluation board allowed for more precise timing and control of all aspects of the illumination, including the relative exposure duration and bit-depth of the red, green and blue channels. For example, in this work  only the red and green channels are used. So for these experiments the projector was programmed to update at 30 Hz and display a green pattern for 235 us (1 bit depth), followed by a red pattern for 33,098 us (8 bit depth) during each 30 Hz cycle. This choice of timing and bit depth maximizes the range of average intensities available in the red channel for optogenetic stimulation, while restricting the green channel to binary images sufficient for calibration. The choice of 30 Hz is optimized for the 30 Hz camera framerate. To avoid aliasing,  camera acquisition was synchronized to the projector  by wiring the camera trigger input to the green LED on the light engine.     If both red and blue channels are to be used for optogenetic stimulation, a different set of timing parameters can be  used.

It is desired that the animal perceives the illumination as continuous and not flickering.  The inactivation time constant for   Chrimson is 21.4$\pm$1.1 ms \cite{klapoetke_independent_2014}. As configured, the  80 uW/mm$^2$ illumination intensity generates a gap in red light illumination of only 235 us from cycle to cycle, well below Chrimson's inactivation time constant. Therefore the animal will perceive the illumination as continuous. At 20 uW/mm$^2$ the  gap in illumination due to the temporal modulation of the micromirrors is nearly 25 ms, similar to  the inactivation timescale.   Intensities lower than this may  be perceived as flickering. The only lower intensities used in this work were 0.5 uW/mm$^2$, for certain control experiments.  We were reassured to observe no obvious behavioral response of any kind to 0.5 uW/mm$^2$ illumination, suggesting that in this case the animal perceived no stimulus at all.

A set of bandpass and longpass filters was used in front of the camera to block  red and blue light from the projector while passing green light for calibration and IR light for behavior imaging. These were, in series, a 538/40 nm bandpass filter (Semrock, FF01-538/40-35-D), a 550 nm longpass filter (Schott, OG-550), and two color photography filters (Roscolux \#318). A 16 mm c-mount lens (Fujinon, CF16ZA-1S)  and spacer (CMSP200, Thorlabs) was used to form an image. 
Barrel distortion is corrected in software.

A PC with a 3.7 GHz CPU  (AMD 3970x) containing 32 cores and a GPU (Quadro P620, Nvidia) controlled the instrument and performed  all real-time processing.  A 6 TB PCIe solid-state drive provided fast writeable storage on which to store high resolution video streams in real-time, Table \ref{tab:bandwidth}. Images from the camera arrived via USB-C.  Drawings were sent to the projector's evaluation board via HDMI.

A complete parts list is provided in \nameref{tab:partslist}, and additional details are described in \cite{liu_c_2020}.

\begin{table}[!ht]
	\begin{adjustwidth}{-2.25in}{0in} 
		\centering
		\caption{
			{\bf Input and output video streams used or generated by the instrument.} Bandwdith is reported in Megabytes per second of the recording.}
		\begin{tabular}{|l|l|l|l|}
			\hline
			Video Stream & Resolution & Format & Bandwidth (MB/s)                                                                                                             \\ \thickhline
			Camera video in (real-time)  & 2048x1504 @ 30Hz & 8-bit monochrome via USB-C & 85.3 
			\\ \hline
			Camera video saved (real-time)  & 2048x1504 @ 30Hz & 8-bit monochrome TIFF & 85.3 
			\\ \hline
			Camera video compressed (post-processing)  & 2048x1504 @ 30Hz & 8-bit monochrome HEVC &  0.102
			\\ \hline
			Projector video out (real-time)  & 912x1140 @ 60Hz & 8-bit RGB* via HDMI & 178 \\ \hline                                                          
			Projector Video Saved (real-time)  & 912x570 @ 30Hz & 8-bit RGB TIFF & 44.5 \\ \hline                                                          
			\begin{tabular}[c]{@{}l@{}}Projector video aligned to camera\\frame of reference (post-processing)\end{tabular}   & 2048x1504 @ 30Hz & 8-bit RGB PNG &  0.353
			\\ \hline      
		\end{tabular}
		\begin{flushleft} *The green channel is actually displayed as binary since it is only used for calibration. Single color experiments in red or blue can achieve 8-bit color resolution but runs at 30 Hz. For experiments with both red and blue, the projector can only simultaneously decode 7-bits of color resolution for each channel but it runs at 60 Hz.  
		\end{flushleft} 
		\label{tab:bandwidth}
	\end{adjustwidth}
\end{table}

\subsubsection*{Real-time software}
\begin{figure}[!htbp]
	\begin{center}
		\includegraphics[width=\textwidth]{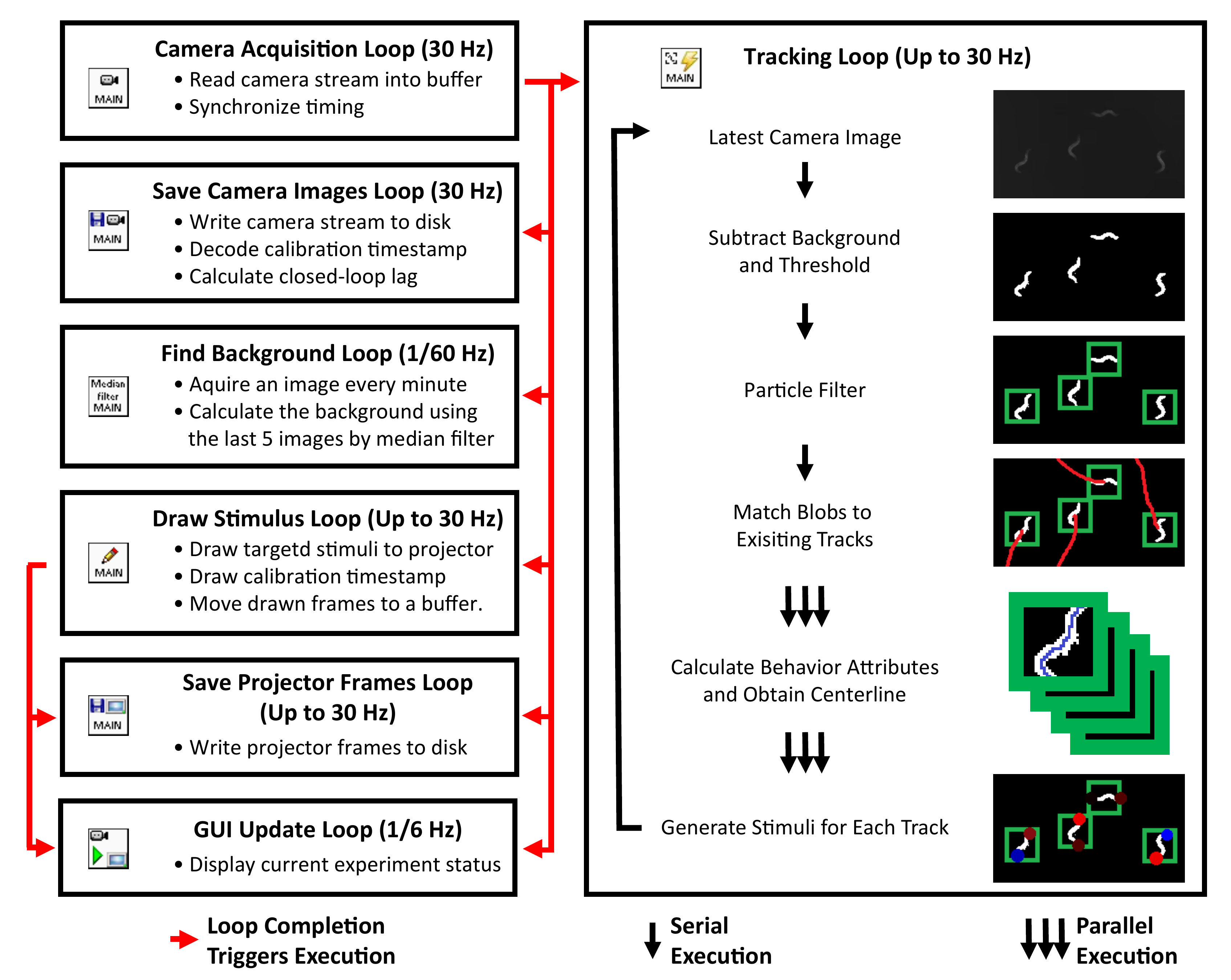}
		\caption{{\bf Selected software modules involved in  closed-loop light stimulation.} Selected software modules are shown that run synchronously or asynchronously. Note the image processing depiction is illustrative and for a small cropped portion of the actual field of view. }
		\label{fig:assaymodules}
	\end{center}
\end{figure}

\begin{figure}[!htbp]

\begin{center}
	\includegraphics[width=1\textwidth]{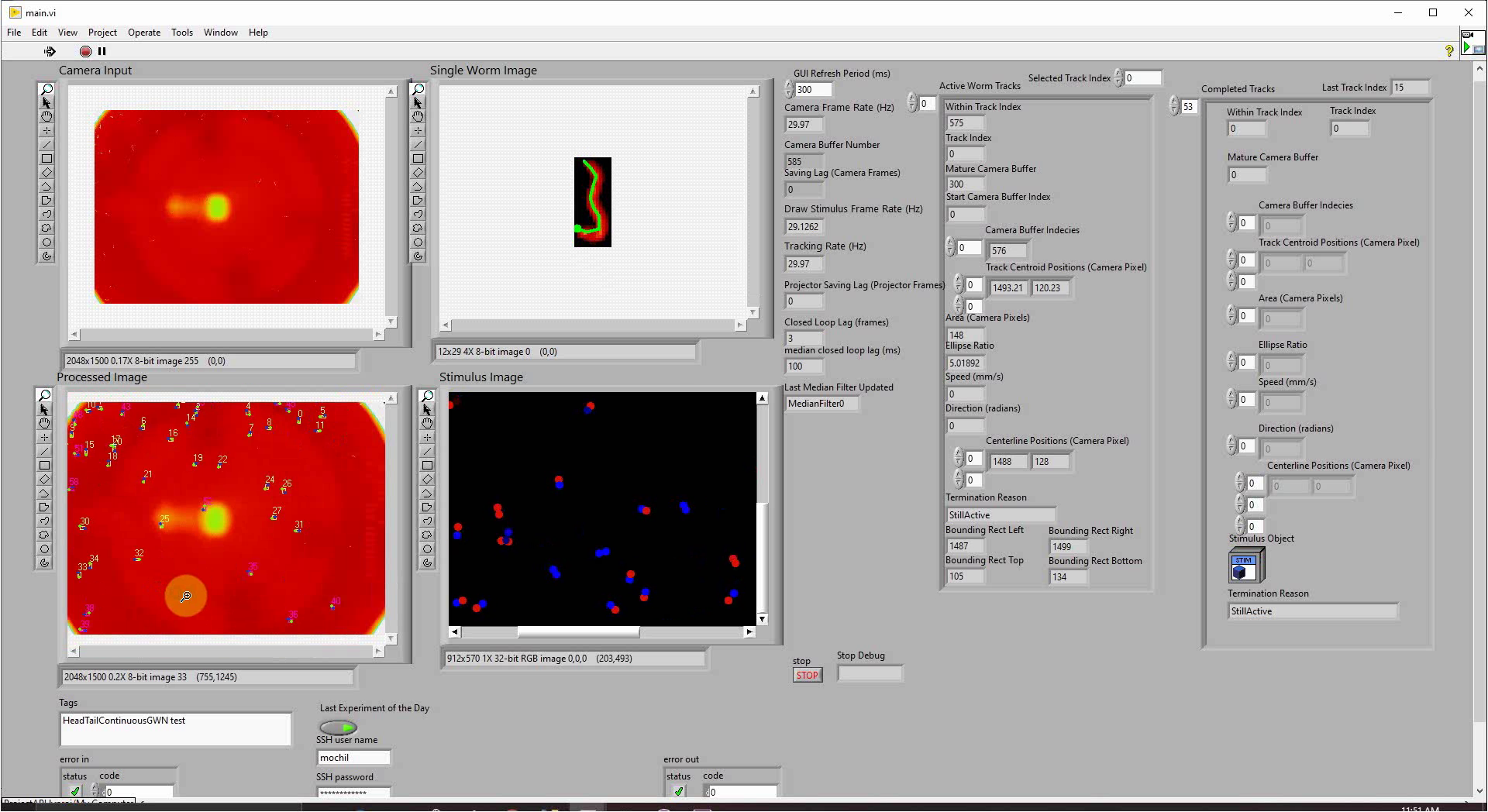}
\end{center}
\caption{{\bf Graphical user interface (GUI)}  shown here during an experiment. }
\label{fig:GUI}
\end{figure}

Custom LabVIEW software  was written to perform all real-time image processing and to control all hardware. Software is available at \url{https://github.com/leiferlab/liu-closed-loop-code}. The LabVIEW software is described in detail in \cite{liu_c_2020}. The software is composed of many modules, summarized in Figure \ref{fig:assaymodules}. These modules  run separately and often asynchronously to acquire images from the camera, track worms, draw new stimuli, communicate with the projector and update a GUI shown in Figure \ref{fig:GUI}. The software  was designed with parallel processing in mind and contains many parallel loops that run independently to take advantage of the multiple cores. 

Camera images and drawn projector images are both saved to disk in real-time  as TIFF images, Table \ref{tab:bandwidth}. In post-processing, they are compressed and converted to H.265 HEVC videos using the GPU.

\begin{figure}[!htbp]
\begin{center}
	\includegraphics[width=1\textwidth]{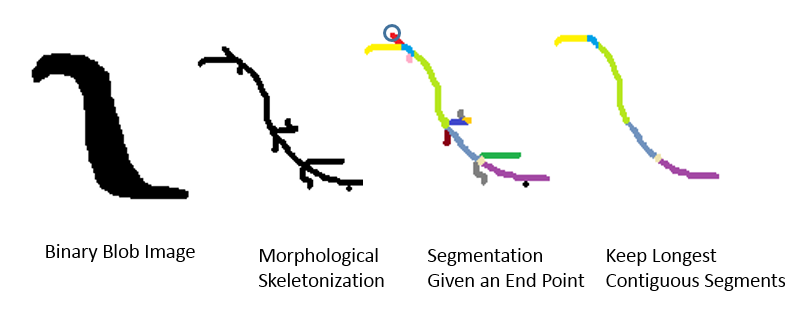}
\end{center}
\caption{{\bf Illustration of the fast centerline finding algorithm.} The algorithm proceeds in order from left to right. A binary image of the worm is taken as input. A skeleton is then generated through morphological thinning. The first recursive algorithm starts from an endpoint of the skeleton and breaks it down into segments at each branch point. The second recursive algorithm uses these segments to find the longest contiguous segment from one end point to another end point.}
\label{fig:centerline}
\end{figure}

A critical task performed by the software is to track each animals' centerline in real-time so that targeting can be delivered to different regions of the animal.  Many centerline algorithms have been developed \cite{husson_keeping_2012}, including some that can operate on a single animal in real-time, e.g. \cite{leifer_optogenetic_2011}. The existing algorithms we tested were too slow to run simultaneously on the many animals needed here. Instead  we developed a two pass recursive algorithm that is fast and computationally efficient, Figure \ref{fig:centerline}.  An image of the worm is binarized and  then skeletonized through morphological thinning. Then, in the first pass, the skeleton is traversed recursively to segment all of the distinct branch points in the skeleton. Then in a second pass, the path of the centerline is found by recursively traversing  all sets of contiguous segments to identify the longest contiguous path. The longest contiguous path is resampled to 20 points and reported as the centerline. Further details are described in \cite{liu_c_2020}.

From the animal's centerline, other key attributes are calculated in real-time, including the animal's velocity and eccentricity which are used to determine whether the animal is turning. The software also stitches together images of worms into tracks, using methods adopted from the Parallel Worm Tracker\cite{ramot_parallel_2008} and described in \cite{liu_temporal_2018} and \cite{liu_c_2020}. To identify the head or tail in real-time, the software assumes that the worm cannot go backwards for more than 10 s, and so the direction of motion for greater than 10 s  indicates the orientation of the head.

\subsubsection*{Registration and Calibration}
Te generate a map between locations in a camera image and locations on the agar, images are acquired of a calibration grid of dots. A transformation is then generated that accounts for barrel distortion and other optical aberrations.  

To generate a map between the projector mirrors and the image viewed by the camera,  the projector draws a spatial calibration pattern in the green channel before each recording. This projector-generated pattern is  segmented in software and automatically creates a mapping between projector and camera image. The calibration is also performed at the end of each recording to quantify any drift that occurred between projector and camera during the course of the recording.  

To provide temporal calibration and to quantify round-trip latency, a visual frame stamp is projected in the green channel for every frame. 
The time-stamp appears as a sequence of dots arranged in a circle, each dot representing one binary digit. Any worms inside the circle are all excluded from analysis. Illumination intensity measurements were taken using a powermeter.  

\subsubsection*{Behavior Analysis}
After images have been compressed in HEVC format, data is sent to Princeton University's high performance computing cluster, Della, for post-processing and data analysis. Custom MATLAB scripts based on \cite{liu_temporal_2018}  inspect and classify animal behavior. For example, the real-time centerline algorithm fails when the animal touches itself. Therefore, when analyzing behavior after the fact, a slower centerline algorithm \cite{deng_efficient_2013}, as implemented in \cite{liu_temporal_2018}, is run to track the centerline even through self-touches. 

The animal's velocity is smoothed with 1 s boxcar window and used to define  behavior states for the experiments in Figures \ref{fig:apresponse} and \ref{fig:heatmap} according to equal area cutoffs shown in \nameref{fig:velocitybehaviors}. 

The turning investigation in Figure \ref{fig:turning} uses the behavior mapping algorithms from \cite{liu_temporal_2018} which in turn are based on an approach first described in \cite{berman_mapping_2014}. The mapping approach classifies the behavior state that the animal occupies at each instance in time. In  many instances the classification system decides that the worm is not in a stereotyped behavior, and therefore it declines to classify the behavior. We inspect even the unclassfied behaviors and classify the animal as turning if a) the behavior mapping algorithm from \cite{liu_temporal_2018} defines it as a turn, or b) if the behavior mapping algorith classifies it as a non-stereotyped behavior that has an eccentricity ratio less than 3.6. In this way we rescue a number of instances of turning that had been overlooked.  For turning experiments in Figure \ref{fig:turning} additional criteria are also used to determine whether a stimulus landed during the turning onset,  and to classify the animal's behavioral response, as described below.

\subsection*{Nematode handling}
Worm preparation was similar to that in \cite{liu_temporal_2018}. To obtain day 2 adults, animals were bleached four days prior to experiment.  To obtain day 1 adults, animals were bleached three days prior to experiments. For optogenetic experiments, bleached worms were placed on plates seeded with 1 ml of 0.5 mM all-trans-retinal (ATR) mixed with OP50 \textit{E. coli}. Off-retinal control plates lacked ATR. Animals were grown in the dark at 20 C.  

To harvest worms for high-throughput experiments, roughly 100 to 200 worms were cut from agar, washed in M9 and then spun-down in a 1.5 ml micro centrifuge tube. For imaging, four small aliquots of worms in M9 were deposited as droplets on the cardinal directions at the edge of the plate. Each droplet typically contained at least  10 worms for a total of approximately 30-50 worms. The droplet was then dried with a tissue.

\subsection*{Anterior vs posterior stimulation experiments}
Day 2 adults were used. Every 30 seconds, each tracked animal was given  a 500 um diameter red light stimulation to either the head, tail, or both   simultaneously. The illumination spot was centered on the tip of the head or tail respectively. The stimulus intensity was randomly drawn  independently for the head and the tail from the set of 0, 20, 40, 60 and 80 uW/mm\textsuperscript{2} intensities.

To calculate the probability of a reversal response for Figure \ref{fig:heatmap}, we first record the animal's behavior 1 second prior to the stimulation to account for any effect the 1 second smoothing window may have on the annotation. Then, for a total of 2 seconds, one second before stimulation and one second during stimulation, we determine if the animal changes behavioral states. The behavioral response   is determined to be the most extreme response the animal has in this 2 second time window as compared to the animal's starting behavior, and it needs to be sustained for more than 0.5 seconds. Note this response may not be the first behavior the animal transitions to, because a paused animal neesd to go through the forward behavior  state before it can enter sprint. 

\subsection*{Mechanosensory evoked response during turning vs forward locomotion }

\subsubsection*{Open-loop whole-body illumination experiments}
Day 1 adults were used.  Every 30 s,  each tracked worm received a 3 s duration 1.5 mm diameter red light illumination spot centered on its body with illumination intensity randomly selected to be either 80 uW/mm\textsuperscript{2} or 0.5 uW/mm\textsuperscript{2}. For a subset of plates  additional illumination intensities were also used, and/or 1 s and 5 s stimulus duration were also used, but those stimulus events were all excluded from analysis in Figure \ref{fig:turning}.  But stimuli of all intensities and duration  were  counted for the purposes of throughput calculations in Table \ref{tab:enrichmentcomparison}, so long as they passed our further criteria for turning onset, worm validity and track stability  described in the next section.

Open-loop stimulation was used to study response to stimulation of animals during forward locomotion. Therefore only stimulus events that landed  when the animal exhibited forward locomotion were included. Forward locomotion was  defined primarily by the behavior mapping algorithm, but we also required   agreement with  our turning onset-detection algorithm, described below.

During post-processing, the following worm stimulation events were excluded from analysis based on track stability and worm validity: Instances in which tracking was lost or the worm exited the field of view 17 seconds before or during the stimulus; instances when the worm collided with another worm before, during, or immediately after stimulation; or stimulations to worms that were stationary, exceedingly fat,  oddly shaped, or were shorter than expected (less than 550 um for these experiments).

\subsubsection*{Closed-loop turn-triggered whole-body illumination experiments}
Whenever the real-time software detected that a tracked worm exhibited the onset of a turn, it delivered 3 s of 1.5mm diameter   red light illumination centered on the worm with an intensity randomly selected to be either 80 uW/mm\textsuperscript{2} or 0.5 uW/mm\textsuperscript{2}. A refractory period was imposed to prevent the same animal from being stimulated twice in less than a 30 s interval. 

Stimulus delivery was triggered by real-time detection of turning onset by triggering on instances when the ellipsoid ratio of  the binarized image of the worm crossed below 3.5.   During post-processing,  a more stringent set of criteria was applied. To be considered a turn onset, the stimulus was required to land also  when the improved behavior mapping pipeline considered the animal to be in a turn. We also required that the stimuli did indeed fall within a 0.33 s window immediately following the ellipse ratio crossing, to account for possible real-time processing errors. We had also observed some instances when tail bends during reversals were incorrectly categorized as turning onset events.  We therefore required that turn onsets occur only when the velocity was above  -0.05 mm/s in a 0.15 s time window prior to stimulus onset. 
Finally, the same exclusion criteria regarding worm validity and tracking stability from the open-loop whole-body illumination experiments were also applied. 

\subsubsection*{Calculating probability of reversals}
To be classified as exhibiting a reversal in response to a stimulation for experiments shown in Figure \ref{fig:turning}, the animal's velocity must decrease below -0.1 mm/s at least once during the 3 s window in which the stimulus is delivered.

\subsection*{Source code}
All software are available at \url{https://github.com/leiferlab/liu-closed-loop-code} including: real-time instrument control software written in LABVIEW, post-processing behavior analysis written in MATLAB, and experiment-specific analysis scripts written in MATLAB.

\subsection*{Datasets}
Datasets for all recordings in this work (\nameref{tab:recordings})   are being deposited in the IEEE DataPorts public repository, DOI:10.21227/t6b0-bc36,   \url{https://dx.doi.org/10.21227/t6b0-bc36}.


\newpage
\section*{Supplementary Figures, Tables and Videos}

\paragraph*{Supplementary Figure S1}
\label{fig:velocitybehaviors}
\begin{center}
	\includegraphics[width=0.8\textwidth]{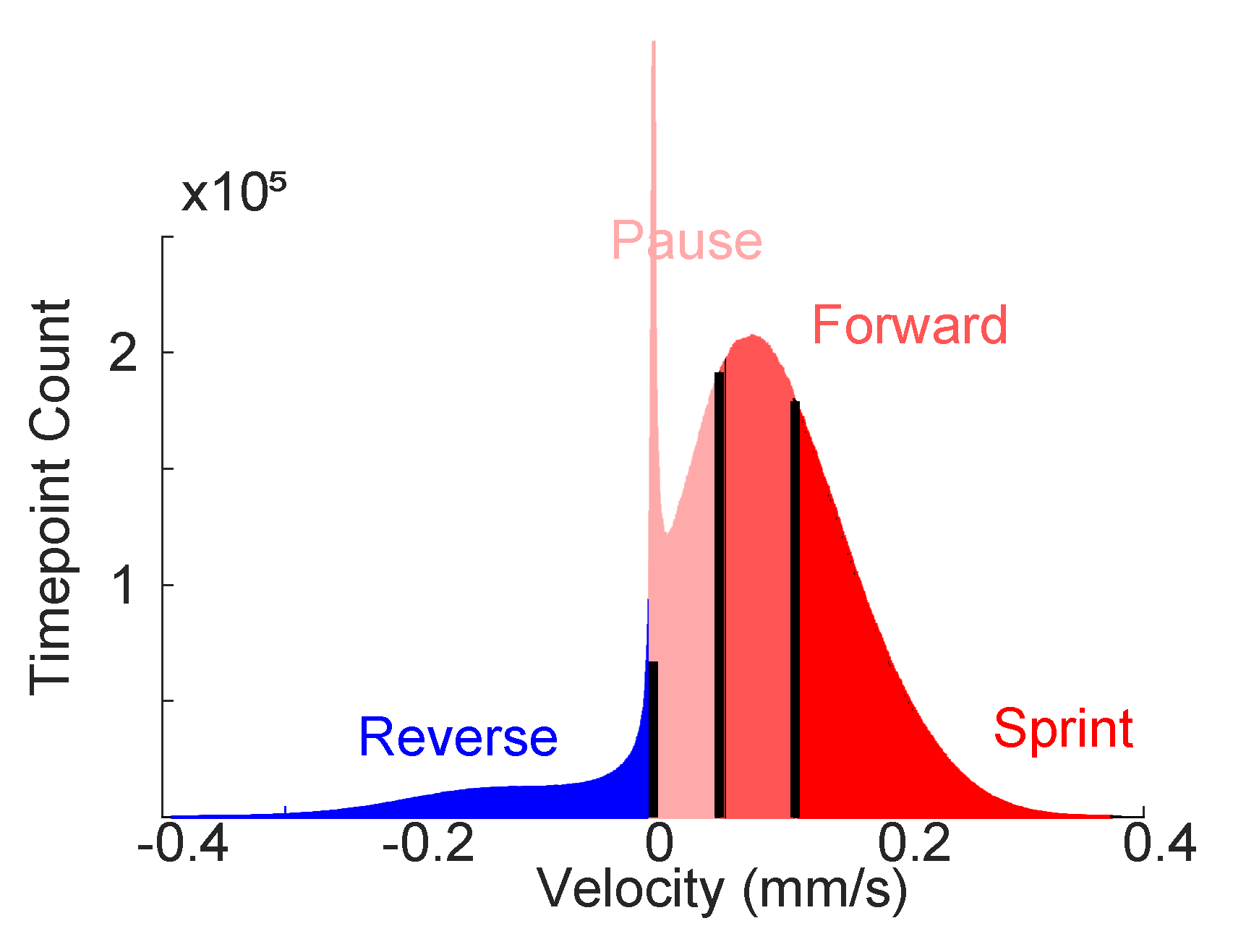} 
\end{center}
{\bf Classifying behavior by velocity distribution.} For the analyses performed in Figures \ref{fig:apresponse} and \ref{fig:heatmap}, we classify animal behavior to be one of four states using the velocity distribution aggregated from both (+)ATR and (-)ATR head/tail stimulation experiments conducted with AML470. The ``Reverse'' state has a velocity of less than 0. The three states with positive velocity are divided so they are equally likely. The slowest state is ``Pause.'' The middle state is ``Forward,''and the fastest state is "Sprint".

\newpage
\paragraph*{Supplementary Figure S2}
\label{fig:anteriorposteriornoret}
\begin{adjustwidth}{-2.25in}{0in}
\begin{center}
	\includegraphics[width=1.45\textwidth]{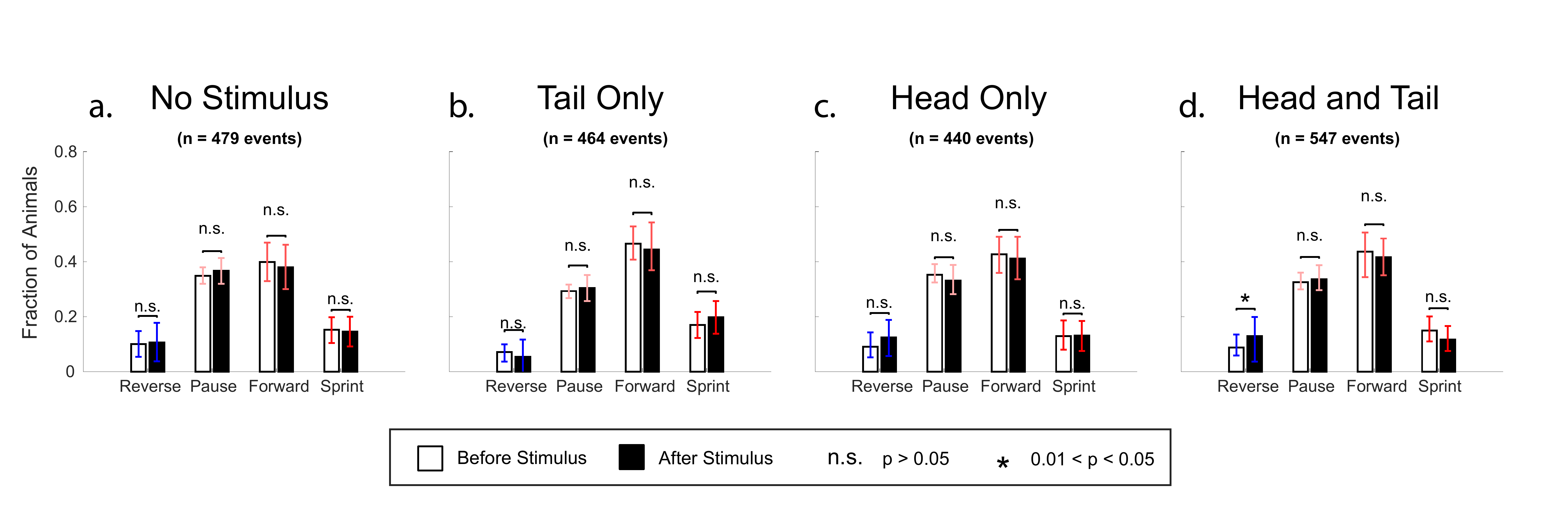} 
\end{center}
{\bf Animals that lack the necessary co-factor all-trans-retinal do not exhibit a behavioral response. } AML470 worms grown (-)ATR. The fraction of animals belonging to behavioral states segmented by velocity before and after each stimulus condition: \textbf{a} no stimulus, \textbf{b} tail only stimulus, \textbf{c} head only stimulus, and \textbf{d} combined head and tail stimulus. The before time point is taken 2 seconds prior to the stimulus onset, and the after time point is taken at the end of the 1 second stimulation. Significance between before and after is determined by p-values calculated using Wilcoxon rank-sum test. The significance test does not correct for multiple hypothesis testing. Error bars represents 95\% confidence intervals estimated using 1000 bootstraps. 
\end{adjustwidth}

\newpage
\paragraph*{Supplementary Figure S3}
\begin{adjustwidth}{-2.25in}{0in}
\label{fig:AML67anteriorposterior}
\begin{center}
	\includegraphics[width=1.4\textwidth]{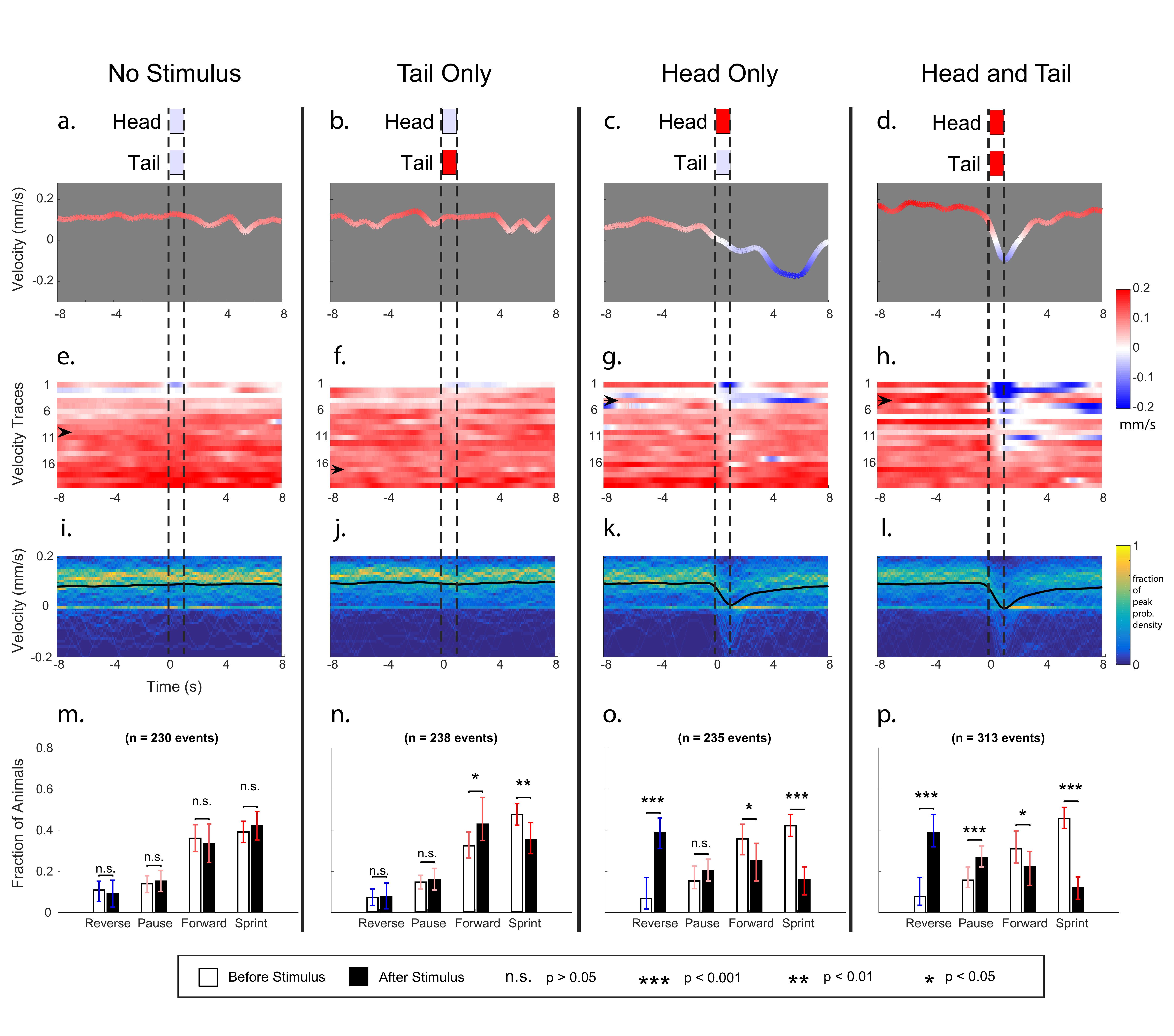}
\end{center}
{\bf Strain AML67 exhibits unexpected response to optogenetic stimulation of posterior mechanosensory neurons}. Anterior targeted stimuli evoke reversals  but posterior stimuli unexpectedly reduce sprints for  worms expressing Chrimson in the soft touch mechanosensory neurons in this strain. Strain AML67 grown on retinal (+)ATR. \textbf{a-d)} Select single animal velocity in response to no stimulation, tail only stimulation, head only stimulation, and combined head and tail stimulation respectively. Dotted lines denote the onset and termination of stimulation. The stimulus is a 1 second long 0.5mm diameter circular dot centered at the tip of the animal's head and/or tail with a red intensity of 80 uW/mm\textsuperscript{2}. \textbf{e-h)} Randomly selected velocity traces for 20 animals in each stimulus condition are color coded red/blue. The traces are sorted by the mean velocity during the 1 second stimulation window. The arrow indicates the selected velocity trace shown in the corresponding column in \textbf{a-d}. \textbf{i-l)} Probability density of all tracked velocities for each stimulus condition. The mean velocity at any given time is overlaid as the black line. The number of stimuli events  for this condition are shown below. \textbf{m-p)} The fraction of animals belonging to behavioral states segmented by velocity before and after each stimulus condition. The before time point is taken 2 seconds prior to the stimulus onset, and the after time point is taken at the end of the 1 second stimulation. Significance between before and after is determined by p-values calculated using Wilcoxon rank-sum test. Error bars represents 95\% confidence intervals estimated using 1000 bootstraps.
\end{adjustwidth}

\newpage
\paragraph*{Supplementary Figure S4}
\label{fig:spectra}
\includegraphics[width=\linewidth]{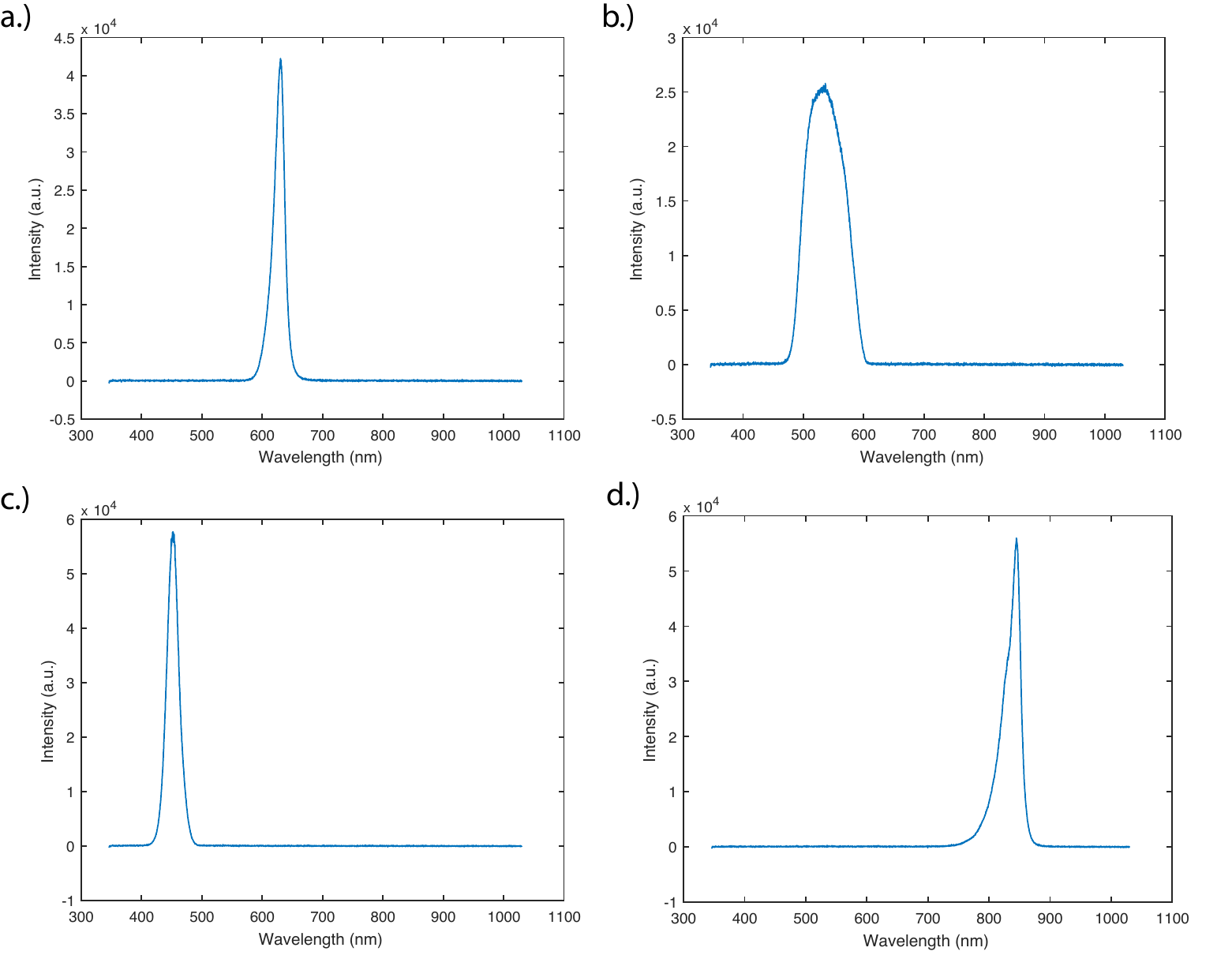}
{\bf Spectral properties of light sources.} Emission spectra are shown for the the projector's a) red b) green and c) blue channels and d) the IR LED ring.

\newpage
\paragraph*{Supplementary Table S1.}
{\bf List of all recordings in this work.} 
\bigskip
\begin{adjustwidth}{-2.25in}{0in}
\label{tab:recordings}
\begin{center}
	\includegraphics[width=1.4\textwidth]{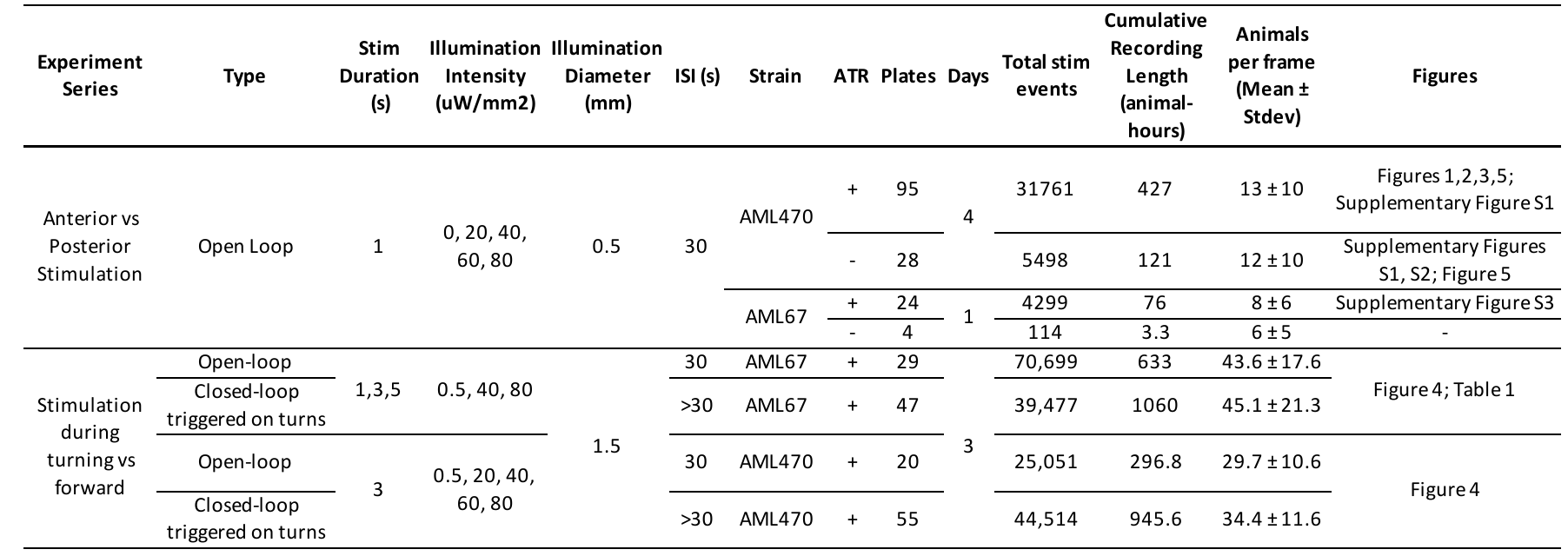}
\end{center}
\end{adjustwidth}
\newpage

\paragraph*{Supplementary Table 2.}
\label{tab:partslist}
{\bf Hardware Partslist.} Hardware parts list for the instrument. The frame is made from 1 inch aluminum extrusions and are not included. The unit count is for each instrument.

\setlength\LTleft{-2.25in}
\setlength\LTright{0in}
\begin{longtable}{lllll}

	\hline
	\multicolumn{1}{|l|}{Manufacturer} &
	\multicolumn{1}{l|}{Item} &
	\multicolumn{1}{l|}{Model/Parts No.} &
	\multicolumn{1}{l|}{Units} &
	\multicolumn{1}{l|}{Notes} \\ \thickhline
	\endfirsthead
	\endhead

	\multicolumn{1}{|l|}{Camera} &
	\multicolumn{1}{l|}{} &
	\multicolumn{1}{l|}{} &
	\multicolumn{1}{l|}{} &
	\multicolumn{1}{l|}{} \\ \hline
	\multicolumn{1}{|l|}{Basler} &
	\multicolumn{1}{l|}{Camera} &
	\multicolumn{1}{l|}{acA4112-30um} &
	\multicolumn{1}{l|}{1} &
	\multicolumn{1}{l|}{} \\ \hline
	\multicolumn{1}{|l|}{Basler} &
	\multicolumn{1}{l|}{USB Cable} &
	\multicolumn{1}{l|}{Basler 2000035994} &
	\multicolumn{1}{l|}{1} &
	\multicolumn{1}{l|}{} \\ \hline
	\multicolumn{1}{|l|}{Basler} &
	\multicolumn{1}{l|}{\begin{tabular}[c]{@{}l@{}}¼-20 Mounting Adapter \\for Basler ace L USB 3.0\end{tabular}} &
	\multicolumn{1}{l|}{Basler 2200000191} &
	\multicolumn{1}{l|}{1} &
	\multicolumn{1}{l|}{} \\ \hline
	\multicolumn{1}{|l|}{Basler} &
	\multicolumn{1}{l|}{\begin{tabular}[c]{@{}l@{}}Digital I/O and power cable, \\6-pin Hirose (Female), 10 m\end{tabular}} &
	\multicolumn{1}{l|}{Basler 2000029411} &
	\multicolumn{1}{l|}{1} &
	\multicolumn{1}{l|}{} \\ \hline
	&
	&
	&
	&
	\\ \hline
	\multicolumn{1}{|l|}{Lens And Filter} &
	\multicolumn{1}{l|}{} &
	\multicolumn{1}{l|}{} &
	\multicolumn{1}{l|}{} &
	\multicolumn{1}{l|}{} \\ \hline
	\multicolumn{1}{|l|}{Fujinon} &
	\multicolumn{1}{l|}{\begin{tabular}[c]{@{}l@{}}CF16ZA-1S 16mm f/1.80\\ Machine Vision C-Mount Lens\end{tabular}} &
	\multicolumn{1}{l|}{CF16ZA-1S} &
	\multicolumn{1}{l|}{1} &
	\multicolumn{1}{l|}{} \\ \hline
	\multicolumn{1}{|l|}{Edmund Optics} &
	\multicolumn{1}{l|}{M37.5 x 0.5 Empty Filter Mount} &
	\multicolumn{1}{l|}{67-684} &
	\multicolumn{1}{l|}{1} &
	\multicolumn{1}{l|}{} \\ \hline
	\multicolumn{1}{|l|}{Thorlabs} &
	\multicolumn{1}{l|}{C-Mount Spacer Ring, 2.00 mm Thick} &
	\multicolumn{1}{l|}{CMSP200} &
	\multicolumn{1}{l|}{1} &
	\multicolumn{1}{l|}{} \\ \hline
	\multicolumn{1}{|l|}{Semrock} &
	\multicolumn{1}{l|}{\begin{tabular}[c]{@{}l@{}}538/40 nm BrightLine® single-band \\bandpass filter, 35mm unmounted\end{tabular}} &
	\multicolumn{1}{l|}{FF01-538/40-35-D} &
	\multicolumn{1}{l|}{1} &
	\multicolumn{1}{l|}{\begin{tabular}[c]{@{}l@{}}Also allows \\in IR \textgreater 800nm\end{tabular}} \\ \hline
	\multicolumn{1}{|l|}{SCHOTT} &
	\multicolumn{1}{l|}{OG-550, 50.8mm Sq., Longpass Filter} &
	\multicolumn{1}{l|}{OG-550} &
	\multicolumn{1}{l|}{1} &
	\multicolumn{1}{l|}{} \\ \hline
	\multicolumn{1}{|l|}{Roscolux} &
	\multicolumn{1}{l|}{\begin{tabular}[c]{@{}l@{}}3" x 5" 200 filters, Color Filter \\Booklet, \#318Mayan Sun\end{tabular}} &
	\multicolumn{1}{l|}{39-418} &
	\multicolumn{1}{l|}{1} &
	\multicolumn{1}{l|}{\begin{tabular}[c]{@{}l@{}}35 mm circles\\ are cut out. \\2 filters are \\used per lens.\end{tabular}} \\ \hline
	&
	&
	&
	&
	\\ \hline
	\multicolumn{1}{|l|}{Projector} &
	\multicolumn{1}{l|}{} &
	\multicolumn{1}{l|}{} &
	\multicolumn{1}{l|}{} &
	\multicolumn{1}{l|}{} \\ \hline
	\multicolumn{1}{|l|}{Anhua} &
	\multicolumn{1}{l|}{\begin{tabular}[c]{@{}l@{}}M5NP Structured light \\projector engine for 3D Measuring\end{tabular}} &
	\multicolumn{1}{l|}{M5NP} &
	\multicolumn{1}{l|}{1} &
	\multicolumn{1}{l|}{} \\ \hline
	\multicolumn{1}{|l|}{Wintech} &
	\multicolumn{1}{l|}{\begin{tabular}[c]{@{}l@{}}DLP4500 Projection DLP Reference \\Design Evaluation Board \\without optical module\end{tabular}} &
	\multicolumn{1}{l|}{LCR4500 VIS KIT} &
	\multicolumn{1}{l|}{1} &
	\multicolumn{1}{l|}{} \\ \hline
	\multicolumn{1}{|l|}{Kaga Electronics USA} &
	\multicolumn{1}{l|}{AC/DC DESKTOP ADAPTER 12V 84W} &
	\multicolumn{1}{l|}{KTPS90-1207} &
	\multicolumn{1}{l|}{1} &
	\multicolumn{1}{l|}{} \\ \hline
	\multicolumn{1}{|l|}{Tripp Lite} &
	\multicolumn{1}{l|}{6' USB A TO MINI-B CABLE M/M} &
	\multicolumn{1}{l|}{UR030-006} &
	\multicolumn{1}{l|}{1} &
	\multicolumn{1}{l|}{} \\ \hline
	\multicolumn{1}{|l|}{Tripp Lite} &
	\multicolumn{1}{l|}{HDMI TO MINI HDMI CABLE 6'} &
	\multicolumn{1}{l|}{P571-006-MINI} &
	\multicolumn{1}{l|}{1} &
	\multicolumn{1}{l|}{} \\ \hline
	\multicolumn{1}{|l|}{CNC Tech} &
	\multicolumn{1}{l|}{\begin{tabular}[c]{@{}l@{}}CORD 18AWG 5-15P - \\320-C5 6' BLK\end{tabular}} &
	\multicolumn{1}{l|}{800-18-39B-BL-0006F} &
	\multicolumn{1}{l|}{1} &
	\multicolumn{1}{l|}{} \\ \hline
	\multicolumn{1}{|l|}{Molex} &
	\multicolumn{1}{l|}{\begin{tabular}[c]{@{}l@{}}Socket Contact Tin \\24-26 AWG Crimp Stamped\end{tabular}} &
	\multicolumn{1}{l|}{874210000} &
	\multicolumn{1}{l|}{50} &
	\multicolumn{1}{l|}{} \\ \hline
	\multicolumn{1}{|l|}{Molex} &
	\multicolumn{1}{l|}{\begin{tabular}[c]{@{}l@{}}9 Position Rectangular Housing \\Connector Receptacle Natural   \\0.059" (1.50mm) for evaluation board\end{tabular}} &
	\multicolumn{1}{l|}{87439-0900} &
	\multicolumn{1}{l|}{1} &
	\multicolumn{1}{l|}{} \\ \hline
	\multicolumn{1}{|l|}{Molex} &
	\multicolumn{1}{l|}{\begin{tabular}[c]{@{}l@{}}6 Position Rectangular Housing \\Connector Receptacle Natural \\0.059" (1.50mm) for evalutaion board\end{tabular}} &
	\multicolumn{1}{l|}{87439-0600} &
	\multicolumn{1}{l|}{3} &
	\multicolumn{1}{l|}{} \\ \hline
	\multicolumn{1}{|l|}{Molex} &
	\multicolumn{1}{l|}{Fan connector} &
	\multicolumn{1}{l|}{51021-0300} &
	\multicolumn{1}{l|}{1} &
	\multicolumn{1}{l|}{} \\ \hline
	\multicolumn{1}{|l|}{Molex} &
	\multicolumn{1}{l|}{Fan and optical module terminal crimp} &
	\multicolumn{1}{l|}{50079-8100} &
	\multicolumn{1}{l|}{20} &
	\multicolumn{1}{l|}{} \\ \hline
	\multicolumn{1}{|l|}{Molex} &
	\multicolumn{1}{l|}{\begin{tabular}[c]{@{}l@{}}6 Position Rectangular Housing \\Connector Receptacle Natural \\0.049" (1.25mm) Optical Module LED\end{tabular}} &
	\multicolumn{1}{l|}{51021-0600} &
	\multicolumn{1}{l|}{3} &
	\multicolumn{1}{l|}{} \\ \hline
	\multicolumn{1}{|l|}{Foxconn} &
	\multicolumn{1}{l|}{Fan} &
	\multicolumn{1}{l|}{PVA030E12M} &
	\multicolumn{1}{l|}{1} &
	\multicolumn{1}{l|}{} \\ \hline
	&
	&
	&
	&
	\\ \hline
	\multicolumn{1}{|l|}{IR Ring} &
	\multicolumn{1}{l|}{} &
	\multicolumn{1}{l|}{} &
	\multicolumn{1}{l|}{} &
	\multicolumn{1}{l|}{} \\ \hline
	\multicolumn{1}{|l|}{LightingWill} &
	\multicolumn{1}{l|}{\begin{tabular}[c]{@{}l@{}}DC12V 5M/16.4ft 72W SMD5050\\ 300LEDs IR InfraRed 850nm Tri-chip White \\PCB Flexible LED Strips 60LEDs 14.4W/M\end{tabular}} &
	\multicolumn{1}{l|}{B01DM9BL5I} &
	\multicolumn{1}{l|}{1} &
	\multicolumn{1}{l|}{} \\ \hline
	\multicolumn{1}{|l|}{DROK} &
	\multicolumn{1}{l|}{\begin{tabular}[c]{@{}l@{}}DC Car Power Supply Voltage \\Regulator Buck Converter 8A/100W \\12A Max DC 5-40V to 1.2-36V Step \\Down Volt Convert Module\end{tabular}} &
	\multicolumn{1}{l|}{90483} &
	\multicolumn{1}{l|}{1} &
	\multicolumn{1}{l|}{} \\ \hline
	\multicolumn{1}{|l|}{CENTROPOWER} &
	\multicolumn{1}{l|}{\begin{tabular}[c]{@{}l@{}}10 Pairs DC Power Pigtail \\Cable 12V 5A Male Female Connectors \\DC Cable for CCTV Security Camera \\Power Adapter Connectors\end{tabular}} &
	\multicolumn{1}{l|}{CT-DCCORD} &
	\multicolumn{1}{l|}{1} &
	\multicolumn{1}{l|}{} \\ \hline
	\multicolumn{1}{|l|}{Marshall} &
	\multicolumn{1}{l|}{\begin{tabular}[c]{@{}l@{}}Potentiometer - 22K Linear,\\ Marshall, 16mm\end{tabular}} &
	\multicolumn{1}{l|}{"0609722158497"} &
	\multicolumn{1}{l|}{1} &
	\multicolumn{1}{l|}{} \\ \hline
	\multicolumn{1}{|l|}{TDK-Lambda Americas} &
	\multicolumn{1}{l|}{AC/DC CONVERTER 24V 27A 600W} &
	\multicolumn{1}{l|}{285-1742-ND} &
	\multicolumn{1}{l|}{1} &
	\multicolumn{1}{l|}{} \\ \hline
	\multicolumn{1}{|l|}{LE PAON} &
	\multicolumn{1}{l|}{\begin{tabular}[c]{@{}l@{}}Embroidery Hoops Plastic \\Cross Stitch Hoop 7 Pcs\end{tabular}} &
	\multicolumn{1}{l|}{4336933944} &
	\multicolumn{1}{l|}{1} &
	\multicolumn{1}{l|}{\begin{tabular}[c]{@{}l@{}}Only the \\6 in hoops \\were used\end{tabular}} \\ \hline
	\multicolumn{1}{|l|}{Thorlabs} &
	\multicolumn{1}{l|}{\begin{tabular}[c]{@{}l@{}}Black Hardboard, 24" x 24" \\(610 mm x 610 mm),   3/16" (4.76 mm) \\Thick, 3 Sheets\end{tabular}} &
	\multicolumn{1}{l|}{TB4} &
	\multicolumn{1}{l|}{1} &
	\multicolumn{1}{l|}{} \\ \hline
	&
	&
	&
	&
	\\ \hline
	\multicolumn{1}{|l|}{Computer} &
	\multicolumn{1}{l|}{} &
	\multicolumn{1}{l|}{} &
	\multicolumn{1}{l|}{} &
	\multicolumn{1}{l|}{} \\ \hline
	\multicolumn{1}{|l|}{Origin PC} &
	\multicolumn{1}{l|}{\begin{tabular}[c]{@{}l@{}}Custom Rackmount PC with \\AMD 3970X Processor\end{tabular}} &
	\multicolumn{1}{l|}{L-Class} &
	\multicolumn{1}{l|}{1} &
	\multicolumn{1}{l|}{} \\ \hline
	\multicolumn{1}{|l|}{Sabrant} &
	\multicolumn{1}{l|}{\begin{tabular}[c]{@{}l@{}}2TB Rocket NVMe PCIe M.2 \\2280 Internal SSD High \\Performance Solid State Drive\end{tabular}} &
	\multicolumn{1}{l|}{SB-ROCKET-2TB} &
	\multicolumn{1}{l|}{4} &
	\multicolumn{1}{l|}{} \\ \hline

\end{longtable}

\newpage
\paragraph*{Supplementary Video S1}
\label{vid:arena}
Video shows excerpt from recording of animals (AML470) crawling on plate and undergoing head and tail stimulation. Yellow numbered `x' indicates a tracked animal, and its track is shown in yellow. Green inset shows a single tracked individual in detail. Green dot indicates the animal's head. Its centerline is shown in green. Tracked animals' heads and tails are occasionally stimulated at various intensities, indicated by red. The dynamic circular pattern in the center of the screen is the visual time stamp projected by the projector and is used for temporal calibration. 

\paragraph*{Supplementary Video S2}
\label{vid:control}
Example responses to control illumination (0 uW/mm$^2$). Strain AML470. The instrument performs all real-time processing to illuminate the head and the tail of each animal for 1 s, but no light is actually delivered to the animal.  Targeted regions are shown in light blue. 20 stimulation-events are shown corresponding to the 20 randomly selected stimulation events shown in Figure \ref{fig:apresponse}e. The animal with the yellow square corresponds to the stimulation event shown in Figure \ref{fig:apresponse}a and denoted with an arrow in Figure  \ref{fig:apresponse}e. The animal's centerline is shown in green. The head of the animal is denoted with a green circle. Note the field of view is centered on each animal, so the animal may be moving even though it never exits the field of view. 

\paragraph*{Supplementary Video S3}
\label{vid:tail}
Example responses to tail illumination (80 uW/mm$^2$).  Posterior mechanosensory neurons are activated by illuminating the tail of animals expression Chrimson int he soft touch mechanosensory neurons (strain AML470). Illumination is indicated by red circles.  20 stimulation-events are shown corresponding to the 20 randomly selected stimulation events shown in Figure \ref{fig:apresponse}f. The animal with the yellow square corresponds to the stimulation event shown in Figure \ref{fig:apresponse}b and denoted with an arrow in Figure \ref{fig:apresponse}f. The animal's centerline is shown in green. The head of the animal is denoted with a green circle. Note the field of view is centered on each animal, so the animal may be moving even though it never exits the field of view. 

\paragraph*{Supplementary Video S4}
\label{vid:head}
Example responses to head illumination (80 uW/mm$^2$).  Similar to \nameref{vid:tail}.  20 stimulation-events are shown corresponding to the 20 randomly selected stimulation events shown in Figure \ref{fig:apresponse}g. The animal with the yellow square corresponds to the stimulation event shown in Figure \ref{fig:apresponse}c.

\paragraph*{Supplementary Video S5}
\label{vid:headandtail}
Example responses to simultaneous head and tail illumination (80 uW/mm$^2$). Similar to \nameref{vid:tail}.  20 stimulation-events are shown corresponding to the 20 randomly selected stimulation events shown in Figure \ref{fig:apresponse}h. The animal with the yellow square corresponds to the stimulation event shown in Figure \ref{fig:apresponse}d.

\section*{Acknowledgments}
Plasmid pCFJ68 was a gift from Erik Jorgensen (University of Utah). We thank Chaogu Zheng (University of Hong Kong) 
for productive discussions.  This work used computing resources from the  Princeton Institute for Computational Science and Engineering. Research reported in this work was supported  by the Simons Foundation under award  SCGB \#543003 to AML;  and  by the National Science Foundation, through an NSF CAREER Award to AML (IOS-1845137) and through the Center for the Physics of Biological Function (PHY-1734030); and by   the National Institute of Neurological Disorders and Stroke of the National Institutes of Health under New Innovator award number DP2-NS116768 to AML. Strains from this work are being distributed by the CGC, which is funded by the NIH Office of Research Infrastructure Programs (P40 OD010440). The content is solely the responsibility of the authors and does not  represent the official views of any funding agency.

\nolinenumbers

%
%
%


\begin{thebibliography}{10}

\bibitem{clark_mapping_2013}
Clark D, Freifeld L, Clandinin T.
\newblock Mapping and {Cracking} {Sensorimotor} {Circuits} in {Genetic} {Model}
  {Organisms}.
\newblock Neuron. 2013;78(4):583--595.
\newblock doi:{10.1016/j.neuron.2013.05.006}.

\bibitem{boyden_history_2011}
Boyden E.
\newblock A history of optogenetics: the development of tools for controlling
  brain circuits with light.
\newblock F1000 Biology Reports. 2011;3.
\newblock doi:{10.3410/B3-11}.

\bibitem{fenno_development_2011}
Fenno L, Yizhar O, Deisseroth K.
\newblock The development and application of optogenetics.
\newblock Annual Review of Neuroscience. 2011;34:389--412.
\newblock doi:{10.1146/annurev-neuro-061010-113817}.

\bibitem{datta_computational_2019}
Datta SR, Anderson DJ, Branson K, Perona P, Leifer A.
\newblock Computational {Neuroethology}: {A} {Call} to {Action}.
\newblock Neuron. 2019;104(1):11--24.
\newblock doi:{10.1016/j.neuron.2019.09.038}.

\bibitem{pereira_quantifying_2020}
Pereira TD, Shaevitz JW, Murthy M.
\newblock Quantifying behavior to understand the brain.
\newblock Nature Neuroscience. 2020;23(12):1537--1549.
\newblock doi:{10.1038/s41593-020-00734-z}.

\bibitem{calhoun_quantifying_2017}
Calhoun AJ, Murthy M.
\newblock Quantifying behavior to solve sensorimotor transformations: advances
  from worms and flies.
\newblock Current Opinion in Neurobiology. 2017;46:90--98.
\newblock doi:{10.1016/j.conb.2017.08.006}.

\bibitem{nagel_light_2005}
Nagel G, Brauner M, Liewald JF, Adeishvili N, Bamberg E, Gottschalk A.
\newblock Light {Activation} of {Channelrhodopsin}-2 in {Excitable} {Cells} of
  {Caenorhabditis} elegans {Triggers} {Rapid} {Behavioral} {Responses}.
\newblock Current Biology. 2005;15(24):2279--2284.
\newblock doi:{10.1016/j.cub.2005.11.032}.

\bibitem{leifer_optogenetics_2012}
Leifer AM.
\newblock Optogenetics and {Computer} {Vision} for {C}. elegans {Neuroscience}
  and {Other} {Biophysical} {Applications} [Thesis].
\newblock Harvard University. Cambridge, MA, USA; 2012.
\newblock Available from:
  \url{http://nrs.harvard.edu/urn-3:HUL.InstRepos:9276708}.

\bibitem{kocabas_controlling_2012}
Kocabas A, Shen CH, Guo ZV, Ramanathan S.
\newblock Controlling interneuron activity in {Caenorhabditis} elegans to evoke
  chemotactic behaviour.
\newblock Nature. 2012;490(7419):273--277.
\newblock doi:{10.1038/nature11431}.

\bibitem{hernandez-nunez_reverse-correlation_2015}
Hernandez-Nunez L, Belina J, Klein M, Si G, Claus L, Carlson JR, et~al.
\newblock Reverse-correlation analysis of navigation dynamics in {Drosophila}
  larva using optogenetics.
\newblock eLife. 2015;4.
\newblock doi:{10.7554/eLife.06225}.

\bibitem{gepner_computations_2015}
Gepner R, Skanata MM, Bernat NM, Kaplow M, Gershow M.
\newblock Computations underlying {Drosophila} photo-taxis, odor-taxis, and
  multi-sensory integration.
\newblock eLife. 2015;4:e06229.
\newblock doi:{10.7554/eLife.06229}.

\bibitem{schulze_dynamical_2015}
Schulze A, Gomez-Marin A, Rajendran VG, Lott G, Musy M, Ahammad P, et~al.
\newblock Dynamical feature extraction at the sensory periphery guides
  chemotaxis.
\newblock eLife. 2015;4:e06694.
\newblock doi:{10.7554/eLife.06694}.

\bibitem{calabrese_search_2015}
Calabrese RL.
\newblock In search of lost scent.
\newblock eLife. 2015;4:e08715.
\newblock doi:{10.7554/eLife.08715}.

\bibitem{gordus_feedback_2015}
Gordus A, Pokala N, Levy S, Flavell SW, Bargmann CI.
\newblock Feedback from network states generates variability in a probabilistic
  olfactory circuit.
\newblock Cell. 2015;161(2):215--227.
\newblock doi:{10.1016/j.cell.2015.02.018}.

\bibitem{claridge-chang_writing_2009}
Claridge-Chang A, Roorda RD, Vrontou E, Sjulson L, Li H, Hirsh J, et~al.
\newblock Writing memories with light-addressable reinforcement circuitry.
\newblock Cell. 2009;139(2):405--415.
\newblock doi:{10.1016/j.cell.2009.08.034}.

\bibitem{cho_parallel_2016}
Cho CE, Brueggemann C, L'Etoile ND, Bargmann CI.
\newblock Parallel encoding of sensory history and behavioral preference during
  {Caenorhabditis} elegans olfactory learning.
\newblock eLife. 2016;5.
\newblock doi:{10.7554/eLife.14000}.

\bibitem{wen_proprioceptive_2012}
Wen Q, Po MD, Hulme E, Chen S, Liu X, Kwok S, et~al.
\newblock Proprioceptive {Coupling} within {Motor} {Neurons} {Drives} {C}.
  elegans {Forward} {Locomotion}.
\newblock Neuron. 2012;76(4):750--761.
\newblock doi:{10.1016/j.neuron.2012.08.039}.

\bibitem{donnelly_monoaminergic_2013}
Donnelly JL, Clark CM, Leifer AM, Pirri JK, Haburcak M, Francis MM, et~al.
\newblock Monoaminergic {Orchestration} of {Motor} {Programs} in a {Complex}
  {C}. elegans {Behavior}.
\newblock PLoS Biology. 2013;11(4):e1001529.
\newblock doi:{10.1371/journal.pbio.1001529}.

\bibitem{kato_global_2015}
Kato S, Kaplan HS, Schrödel T, Skora S, Lindsay TH, Yemini E, et~al.
\newblock Global brain dynamics embed the motor command sequence of
  {Caenorhabditis} elegans.
\newblock Cell. 2015;163(3):656--669.
\newblock doi:{10.1016/j.cell.2015.09.034}.

\bibitem{wang_flexible_2020}
Wang Y, Zhang X, Xin Q, Hung W, Florman J, Huo J, et~al.
\newblock Flexible motor sequence generation during stereotyped escape
  responses.
\newblock eLife. 2020;9:e56942.
\newblock doi:{10.7554/eLife.56942}.

\bibitem{cande_optogenetic_2018}
Cande J, Namiki S, Qiu J, Korff W, Card GM, Shaevitz JW, et~al.
\newblock Optogenetic dissection of descending behavioral control in
  {Drosophila}.
\newblock eLife. 2018;7:e34275.
\newblock doi:{10.7554/eLife.34275}.

\bibitem{boulin_reporter_2006}
Boulin T.
\newblock Reporter gene fusions.
\newblock In: elegans Research~Community TC, editor. {WormBook}; 2006.Available
  from:
  \url{http://www.wormbook.org/chapters/www_reportergenefusions/reportergenefusions.html}.

\bibitem{pfeiffer_tools_2008}
Pfeiffer BD, Jenett A, Hammonds AS, Ngo TTB, Misra S, Murphy C, et~al.
\newblock Tools for neuroanatomy and neurogenetics in {Drosophila}.
\newblock Proceedings of the National Academy of Sciences.
  2008;105(28):9715--9720.
\newblock doi:{10.1073/pnas.0803697105}.

\bibitem{jenett_gal4-driver_2012}
Jenett A, Rubin GM, Ngo TTB, Shepherd D, Murphy C, Dionne H, et~al.
\newblock A {GAL4}-{Driver} {Line} {Resource} for {Drosophila} {Neurobiology}.
\newblock Cell Reports. 2012;2(4):991--1001.
\newblock doi:{10.1016/j.celrep.2012.09.011}.

\bibitem{guo_optical_2009}
Guo ZV, Hart AC, Ramanathan S.
\newblock Optical interrogation of neural circuits in {Caenorhabditis} elegans.
\newblock Nature methods. 2009;6(12):891--896.
\newblock doi:{10.1038/nmeth.1397}.

\bibitem{wyart_optogenetic_2009}
Wyart C, Del~Bene F, Warp E, Scott EK, Trauner D, Baier H, et~al.
\newblock Optogenetic dissection of a behavioural module in the vertebrate
  spinal cord.
\newblock Nature. 2009;461(7262):407--410.
\newblock doi:{10.1038/nature08323}.

\bibitem{leifer_optogenetic_2011}
Leifer AM, Fang-Yen C, Gershow M, Alkema MJ, Samuel ADT.
\newblock Optogenetic manipulation of neural activity in freely moving
  {Caenorhabditis} elegans.
\newblock Nature Methods. 2011;8(2):147--152.
\newblock doi:{10.1038/nmeth.1554}.

\bibitem{stirman_real-time_2011}
Stirman JN, Crane MM, Husson SJ, Wabnig S, Schultheis C, Gottschalk A, et~al.
\newblock Real-time multimodal optical control of neurons and muscles in freely
  behaving {Caenorhabditis} elegans.
\newblock Nature methods. 2011;8(2):153--158.
\newblock doi:{10.1038/nmeth.1555}.

\bibitem{bath_flymad_2014}
Bath DE, Stowers JR, Hörmann D, Poehlmann A, Dickson BJ, Straw AD.
\newblock {FlyMAD}: rapid thermogenetic control of neuronal activity in freely
  walking {Drosophila}.
\newblock Nature Methods. 2014;11(7):756--762.
\newblock doi:{10.1038/nmeth.2973}.

\bibitem{shipley_simultaneous_2014}
Shipley FB, Clark CM, Alkema MJ, Leifer AM.
\newblock Simultaneous optogenetic manipulation and calcium imaging in freely
  moving {C}. elegans.
\newblock Frontiers in Neural Circuits. 2014;8.
\newblock doi:{10.3389/fncir.2014.00028}.

\bibitem{porto_reverse-correlation_2017}
Porto DA, Giblin J, Zhao Y, Lu H.
\newblock Reverse-{Correlation} {Analysis} of {Mechanosensation} {Circuit} in
  {C}. elegans {Reveals} {Temporal} and {Spatial} {Encoding}.
\newblock bioRxiv. 2017;doi:{10.1101/147363}.

\bibitem{dong_toward_2021}
Dong X, Kheiri S, Lu Y, Xu Z, Zhen M, Liu X.
\newblock Toward a living soft microrobot through optogenetic locomotion
  control of {Caenorhabditis} elegans.
\newblock Science Robotics. 2021;6(55).
\newblock doi:{10.1126/scirobotics.abe3950}.

\bibitem{stephens_dimensionality_2008}
Stephens GJ, Johnson-Kerner B, Bialek W, Ryu WS.
\newblock Dimensionality and {Dynamics} in the {Behavior} of {C}. elegans.
\newblock PLoS Computational Biology. 2008;4(4):e1000028.
\newblock doi:{10.1371/journal.pcbi.1000028}.

\bibitem{faumont_image-free_2011}
Faumont S, Rondeau G, Thiele TR, Lawton KJ, McCormick KE, Sottile M, et~al.
\newblock An {Image}-{Free} {Opto}-{Mechanical} {System} for {Creating}
  {Virtual} {Environments} and {Imaging} {Neuronal} {Activity} in {Freely}
  {Moving} {Caenorhabditis} elegans.
\newblock PLoS ONE. 2011;6(9):e24666.
\newblock doi:{10.1371/journal.pone.0024666}.

\bibitem{musso_closed-loop_2019}
Musso PY, Junca P, Jelen M, Feldman-Kiss D, Zhang H, Chan RC, et~al.
\newblock Closed-loop optogenetic activation of peripheral or central neurons
  modulates feeding in freely moving {Drosophila}.
\newblock eLife. 2019;8:e45636.
\newblock doi:{10.7554/eLife.45636}.

\bibitem{adamantidis_optogenetic_2011}
Adamantidis AR, Tsai HC, Boutrel B, Zhang F, Stuber GD, Budygin EA, et~al.
\newblock Optogenetic {Interrogation} of {Dopaminergic} {Modulation} of the
  {Multiple} {Phases} of {Reward}-{Seeking} {Behavior}.
\newblock Journal of Neuroscience. 2011;31(30):10829--10835.
\newblock doi:{10.1523/JNEUROSCI.2246-11.2011}.

\bibitem{oconnor_neural_2013}
O'Connor DH, Hires SA, Guo ZV, Li N, Yu J, Sun QQ, et~al.
\newblock Neural coding during active somatosensation revealed using illusory
  touch.
\newblock Nature Neuroscience. 2013;16(7):958--965.
\newblock doi:{10.1038/nn.3419}.

\bibitem{clancy_volitional_2014}
Clancy KB, Koralek AC, Costa RM, Feldman DE, Carmena JM.
\newblock Volitional modulation of optically recorded calcium signals during
  neuroprosthetic learning.
\newblock Nature Neuroscience. 2014;17(6):807--809.
\newblock doi:{10.1038/nn.3712}.

\bibitem{grosenick_closed-loop_2015}
Grosenick L, Marshel JH, Deisseroth K.
\newblock Closed-loop and activity-guided optogenetic control.
\newblock Neuron. 2015;86(1):106--139.
\newblock doi:{10.1016/j.neuron.2015.03.034}.

\bibitem{krakauer_neuroscience_2017}
Krakauer JW, Ghazanfar AA, Gomez-Marin A, MacIver MA, Poeppel D.
\newblock Neuroscience {Needs} {Behavior}: {Correcting} a {Reductionist}
  {Bias}.
\newblock Neuron. 2017;93(3):480--490.
\newblock doi:{10.1016/j.neuron.2016.12.041}.

\bibitem{stirman_high-throughput_2010}
Stirman JN, Brauner M, Gottschalk A, Lu H.
\newblock High-throughput study of synaptic transmission at the neuromuscular
  junction enabled by optogenetics and microfluidics.
\newblock Journal of Neuroscience Methods. 2010;191(1):90--93.
\newblock doi:{10.1016/j.jneumeth.2010.05.019}.

\bibitem{lee_compressed_2019}
Lee JB, Yonar A, Hallacy T, Shen CH, Milloz J, Srinivasan J, et~al.
\newblock A compressed sensing framework for efficient dissection of neural
  circuits.
\newblock Nature Methods. 2019;16(1):126.
\newblock doi:{10.1038/s41592-018-0233-6}.

\bibitem{wu_optogenetic_2014}
Wu MC, Chu LA, Hsiao PY, Lin YY, Chi CC, Liu TH, et~al.
\newblock Optogenetic control of selective neural activity in multiple freely
  moving {Drosophila} adults.
\newblock Proceedings of the National Academy of Sciences.
  2014;111(14):5367--5372.
\newblock doi:{10.1073/pnas.1400997111}.

\bibitem{kerr_imaging_2006}
Kerr R.
\newblock Imaging the activity of neurons and muscles.
\newblock WormBook. 2006;doi:{10.1895/wormbook.1.113.1}.

\bibitem{branson_high-throughput_2009}
Branson K, Robie AA, Bender J, Perona P, Dickinson MH.
\newblock High-throughput ethomics in large groups of {Drosophila}.
\newblock Nature Methods. 2009;6(6):451--457.
\newblock doi:{10.1038/nmeth.1328}.

\bibitem{gershow_controlling_2012}
Gershow M, Berck M, Mathew D, Luo L, Kane EA, Carlson JR, et~al.
\newblock Controlling airborne cues to study small animal navigation.
\newblock Nature Methods. 2012;9(3):290--296.
\newblock doi:{10.1038/nmeth.1853}.

\bibitem{husson_keeping_2012}
Husson SJ.
\newblock Keeping track of worm trackers.
\newblock In: elegans Research~Community TC, editor. {WormBook}. {WormBook};
  2012.Available from: \url{http://www.wormbook.org}.

\bibitem{liu_temporal_2018}
Liu M, Sharma AK, Shaevitz JW, Leifer AM.
\newblock Temporal processing and context dependency in {Caenorhabditis}
  elegans response to mechanosensation.
\newblock eLife. 2018;7:e36419.
\newblock doi:{10.7554/eLife.36419}.

\bibitem{deangelis_spatiotemporally_2020}
DeAngelis BD, Zavatone-Veth JA, Gonzalez-Suarez AD, Clark DA.
\newblock Spatiotemporally precise optogenetic activation of sensory neurons in
  freely walking {Drosophila}.
\newblock eLife. 2020;9:e54183.
\newblock doi:{10.7554/eLife.54183}.

\bibitem{meloni_controlling_2020}
Meloni I, Sachidanandan D, Thum AS, Kittel RJ, Murawski C.
\newblock Controlling the behaviour of {Drosophila} melanogaster via smartphone
  optogenetics.
\newblock Scientific Reports. 2020;10(1):17614.
\newblock doi:{10.1038/s41598-020-74448-4}.

\bibitem{chalfie_developmental_1981}
Chalfie M, Sulston J.
\newblock Developmental genetics of the mechanosensory neurons of
  {Caenorhabditis} elegans.
\newblock Developmental Biology. 1981;82(2):358--370.

\bibitem{chalfie_neural_1985}
Chalfie M, Sulston JE, White JG, Southgate E, Thomson JN, Brenner S.
\newblock The neural circuit for touch sensitivity in {Caenorhabditis} elegans.
\newblock The Journal of Neuroscience: The Official Journal of the Society for
  Neuroscience. 1985;5(4):956--64.
\newblock doi:{3981252}.

\bibitem{chiba_developmental_1990}
Chiba CM, Rankin CH.
\newblock A developmental analysis of spontaneous and reflexive reversals in
  the {nematodeCaenorhabditis} elegans.
\newblock Journal of Neurobiology. 1990;21(4):543--554.
\newblock doi:{10.1002/neu.480210403}.

\bibitem{chalfie_assaying_2014}
Chalfie M, Hart AC, Rankin CH, Goodman MB.
\newblock Assaying mechanosensation.
\newblock In: elegans Research~Community TC, editor. {WormBook}; 2014.Available
  from: \url{http://doi.org/10.1895/wormbook}.

\bibitem{wicks_integration_1995}
Wicks SR, Rankin CH.
\newblock Integration of mechanosensory stimuli in {Caenorhabditis} elegans.
\newblock The Journal of Neuroscience: The Official Journal of the Society for
  Neuroscience. 1995;15(3 Pt 2):2434--2444.

\bibitem{petzold_mems-based_2013}
Petzold BC, Park SJ, Mazzochette EA, Goodman MB, Pruitt BL.
\newblock {MEMS}-based force-clamp analysis of the role of body stiffness in
  {C}. elegans touch sensation.
\newblock Integrative Biology. 2013;5(6):853--864.
\newblock doi:{10.1039/c3ib20293c}.

\bibitem{mazzochette_real_2013}
Mazzochette EA, Fang-Yen C, Goodman MB, Pruitt BL.
\newblock A {Real} {Time} {Imaging} {System} for {Tracking} {Freely} {Moving}
  {C}. elegans in {Touch} {Assays}.
\newblock In: Microtechnologies in {Medicine} and {Biology}. Marina del Rey,
  CA; 2013.Available from:
  \url{http://microsystems.stanford.edu/microwiki_upload/4/4b/Mazzochette_MMB.pdf}.

\bibitem{mazzochette_tactile_2018}
Mazzochette EA, Nekimken AL, Loizeau F, Whitworth J, Huynh B, Goodman MB,
  et~al.
\newblock The tactile receptive fields of freely moving {Caenorhabditis}
  elegans nematodes.
\newblock Integrative Biology. 2018;10(8):450--463.
\newblock doi:{10.1039/c8ib00045j}.

\bibitem{mcclanahan_comparing_2017}
McClanahan PD, Xu JH, Fang-Yen C.
\newblock Comparing {Caenorhabditis} elegans gentle and harsh touch response
  behavior using a multiplexed hydraulic microfluidic device.
\newblock Integrative Biology. 2017;9(10):800--809.
\newblock doi:{10.1039/C7IB00120G}.

\bibitem{chalfie_neurosensory_2009}
Chalfie M.
\newblock Neurosensory mechanotransduction.
\newblock Nature Reviews Molecular Cell Biology. 2009;10(1):44--52.
\newblock doi:{10.1038/nrm2595}.

\bibitem{gray_circuit_2005}
Gray JM, Hill JJ, Bargmann CI.
\newblock A circuit for navigation in {Caenorhabditis} elegans.
\newblock Proceedings of the National Academy of Sciences of the United States
  of America. 2005;102(9):3184--3191.
\newblock doi:{10.1073/pnas.0409009101}.

\bibitem{alkema_tyramine_2005}
Alkema MJ, Hunter-Ensor M, Ringstad N, Horvitz HR.
\newblock Tyramine {Functions} independently of octopamine in the
  {Caenorhabditis} elegans nervous system.
\newblock Neuron. 2005;46(2):247--260.
\newblock doi:{10.1016/j.neuron.2005.02.024}.

\bibitem{swierczek_high-throughput_2011}
Swierczek NA, Giles AC, Rankin CH, Kerr RA.
\newblock High-throughput behavioral analysis in {C}. elegans.
\newblock Nature Methods. 2011;8(7):592--598.
\newblock doi:{10.1038/nmeth.1625}.

\bibitem{pierce-shimomura_fundamental_1999}
Pierce-Shimomura JT, Morse TM, Lockery SR.
\newblock The {Fundamental} {Role} of {Pirouettes} in {Caenorhabditis} elegans
  {Chemotaxis}.
\newblock J Neurosci. 1999;19(21):9557--9569.

\bibitem{pirri_neuroethology_2012}
Pirri JK, Alkema MJ.
\newblock The neuroethology of {C}. elegans escape.
\newblock Current Opinion in Neurobiology. 2012;22(2):187--193.
\newblock doi:{10.1016/j.conb.2011.12.007}.

\bibitem{maguire_c._2011}
Maguire SM, Clark CM, Nunnari J, Pirri JK, Alkema MJ.
\newblock The {C}. elegans touch response facilitates escape from predacious
  fungi.
\newblock Current biology: CB. 2011;21(15):1326--1330.
\newblock doi:{10.1016/j.cub.2011.06.063}.

\bibitem{govorunova_natural_2015}
Govorunova EG, Sineshchekov OA, Janz R, Liu X, Spudich JL.
\newblock Natural light-gated anion channels: {A} family of microbial
  rhodopsins for advanced optogenetics.
\newblock Science. 2015;349(6248):647--650.
\newblock doi:{10.1126/science.aaa7484}.

\bibitem{vierock_bipoles_2021}
Vierock J, Rodriguez-Rozada S, Dieter A, Pieper F, Sims R, Tenedini F, et~al.
\newblock {BiPOLES} is an optogenetic tool developed for bidirectional
  dual-color control of neurons.
\newblock Nature Communications. 2021;12(1):4527.
\newblock doi:{10.1038/s41467-021-24759-5}.

\bibitem{croll_behavoural_1975}
Croll N.
\newblock Behavoural analysis of nematode movement.
\newblock Advances in Parasitology. 1975;13:71--122.

\bibitem{croll_components_1975}
Croll NA.
\newblock Components and patterns in the behaviour of the nematode
  {Caenorhabditis} elegans.
\newblock Journal of Zoology. 1975;176(2):159--176.
\newblock doi:{10.1111/j.1469-7998.1975.tb03191.x}.

\bibitem{kaplan_nested_2020}
Kaplan HS, Salazar~Thula O, Khoss N, Zimmer M.
\newblock Nested {Neuronal} {Dynamics} {Orchestrate} a {Behavioral} {Hierarchy}
  across {Timescales}.
\newblock Neuron. 2020;105(3):562--576.e9.
\newblock doi:{10.1016/j.neuron.2019.10.037}.

\bibitem{noma_rapid_2018}
Noma K, Jin Y.
\newblock Rapid {Integration} of {Multi}-copy {Transgenes} {Using}
  {Optogenetic} {Mutagenesis} in {Caenorhabditis} elegans.
\newblock G3 Genes{\textbar}Genomes{\textbar}Genetics. 2018;8(6):2091--2097.
\newblock doi:{10.1534/g3.118.200158}.

\bibitem{klapoetke_independent_2014}
Klapoetke NC, Murata Y, Kim SS, Pulver SR, Birdsey-Benson A, Cho YK, et~al.
\newblock Independent optical excitation of distinct neural populations.
\newblock Nature methods. 2014;11(3):338--346.
\newblock doi:{10.1038/nmeth.2836}.

\bibitem{liu_c_2020}
Liu M.
\newblock C. elegans behaviors and their mechanosensory drivers.
\newblock Princeton University. Princeton, United States of America; 2020.
\newblock Available from:
  \url{http://arks.princeton.edu/ark:/88435/dsp01tt44pq78z}.

\bibitem{ramot_parallel_2008}
Ramot D, Johnson BE, Berry TL Jr, Carnell L, Goodman MB.
\newblock The {Parallel} {Worm} {Tracker}: {A} {Platform} for {Measuring}
  {Average} {Speed} and {Drug}-{Induced} {Paralysis} in {Nematodes}.
\newblock PLoS ONE. 2008;3(5):e2208.
\newblock doi:{10.1371/journal.pone.0002208}.

\bibitem{deng_efficient_2013}
Deng Y, Coen P, Sun M, Shaevitz JW.
\newblock Efficient {Multiple} {Object} {Tracking} {Using} {Mutually}
  {Repulsive} {Active} {Membranes}.
\newblock PLoS ONE. 2013;8(6):e65769.
\newblock doi:{10.1371/journal.pone.0065769}.

\bibitem{berman_mapping_2014}
Berman GJ, Choi DM, Bialek W, Shaevitz JW.
\newblock Mapping the stereotyped behaviour of freely moving fruit flies.
\newblock Journal of the Royal Society, Interface / the Royal Society.
  2014;11(99).
\newblock doi:{10.1098/rsif.2014.0672}.

\end{thebibliography}
\end{document}